\documentclass[12pt]{article}
\usepackage{a4wide}
\usepackage{psfrag}
\usepackage{graphics}
\usepackage{amsfonts}
\usepackage{amstext}
\usepackage{amsmath}
\usepackage{amssymb}
\usepackage{amsthm}
\usepackage[mathscr]{eucal}
\usepackage{cite}
\newtheorem{theorem}{Theorem}[section]
\newtheorem{proposition}[theorem]{Proposition}
\def\a{\alpha}

\def\Ai{{\rm Ai}}
\def\b{\beta}
\def\D{\mathcal{D}}
\def\Dg{\mathbb{D}}

\def\g{\gamma}
\def\G{\mathcal{G}}
\def\i{\infty}
\def\I{\mathcal{I}}
\def\Ig{\mathbb{I}}
\def\K{\mathcal{K}}
\def\l{\lambda}
\def\N{\mathbb{N}}
\def\o{\omega}
\def\p{\phi}
\def\vp{\varphi}
\def\P{\mathbb{P}}
\def\R{\mathbb{R}}
\def\sgn{\text{sgn}}

\def\S{\mathcal{S}}
\def\Sg{\mathbb{S}}
\def\t{\tau}

\def\fz{\frac{1}{z_1}}
\def\fw{\frac{1}{z_2}}
\def\Z{\mathbb{Z}}
\def\z{\zeta}

\begin{document}       
\title{Fluctuations of a one-dimensional polynuclear growth model 
in a half space}

%\author{
%\renewcommand{\thefootnote}{\arabic{footnote}}
%\vspace{5mm}
%T. Sasamoto \footnotemark[1] ~and T. Imamura \footnotemark[2]
%}
%\footnotetext[1] 
%{Department of Physics, Tokyo Institute of Technology,
%Oh-okayama 2-12-1, Meguro-ku, Tokyo 152-\\\qquad 8551, Japan.
%e-mail: sasamoto@stat.phys.titech.ac.jp
%tel: +81-3-5734-2727 fax: +81-3-5734-2739 % for JSP
%}
%\footnotetext[2]
%{Department of Physics, Graduate School of Science,
%University of Tokyo,
%Hongo 7-3-1, Bunkyo-ku, \\\qquad Tokyo 113-0033, Japan.
%e-mail: imamura@monet.phys.s.u-tokyo.ac.jp}

\author{
\vspace{5mm}
T. Sasamoto 
{\footnote {\tt e-mail: sasamoto@stat.phys.titech.ac.jp}}
~~and T. Imamura 
{\footnote {\tt e-mail: imamura@monet.phys.s.u-tokyo.ac.jp}}
\\
{\it Department of Physics, Tokyo Institute of Technology,}\\
\vspace{5mm}
{\it Oh-okayama 2-12-1, Meguro-ku, Tokyo 152-8551, Japan}\\
{\it Department of Physics, Graduate School of Science,}\\
{\it University of Tokyo,}\\
{\it Hongo 7-3-1, Bunkyo-ku, Tokyo 113-0033, Japan}\\
}

\date{} 
\maketitle

\begin{abstract}
We consider the multi-point equal time height fluctuations 
of a one-dimensional polynuclear growth model in a half space.
For special values of the nucleation rate at the origin, 
the multi-layer version of the model is reduced to a determinantal 
process, for which the asymptotics can be analyzed.
In the scaling limit, the fluctuations near the origin 
are shown to be equivalent to those of the largest eigenvalue 
of the orthogonal/symplectic to unitary transition ensemble at 
soft edge in random matrix theory.

\vspace{3mm}\noindent
[Keywords: polynuclear growth; KPZ universality class;  
random matrices; Tracy-Widom distribution; Airy process]

%\vspace{5mm}\noindent
%[suggested running head: 1D PNG Model in a Half Space ]

\end{abstract}

\newpage
\section{Introduction}
\label{intro}
The problems of random surface growth have been an 
important subject of non-equilibrium statistical physics
\cite{KS1992,Me1998}.
Interestingly, surfaces show universal statistical behaviors,
depending on which mechanism plays a major role in growth.
Among them is the Kardar-Parisi-Zhang (KPZ) universality 
class \cite{KPZ1986}.
Though in many cases it is not enough to describe the growth 
of real materials, the KPZ universality class gives a satisfactory
understanding of many growth models such as the Eden model, and 
hence plays a prominent role in theoretical study of 
growing surfaces.

It is difficult to study the KPZ universality 
class analytically for general dimension. In one spatial dimension, 
however, some exact results had been obtained. 
For instance, the exponents are already obtained 
in \cite{KPZ1986} based on dynamical renormalization group techniques. 
These have been further confirmed by exact solutions for the 
models in the KPZ universality class \cite{GS1992,Ki1995}.
We will also restrict our attention to one spatial dimensional 
case in this article.

Recently, more refined information about the fluctuation 
properties of the 1+1 dimensional KPZ universality class 
have been obtained 
\cite{Jo2000,PS2000a,PS2000b,BR2000,GTW2001,GTW2002a,GTW2002b,
PS2002a,PS2002b,PS2002p}. 
For some important quantities such as 
the height fluctuation, not only the exponents but also the 
scaling functions are obtained.
The whole new developments originate from the work 
\cite{BDJ1999} on the statistics of the longest increasing 
subsequence in a random permutation, which has turned out 
to be closely related to the random surface growth models
\cite{AD1999}.

The polynuclear growth (PNG) models are known to belong to 
the KPZ universality class \cite{KS1992,Me1998}.
The PNG models are simple models which describe layer by 
layer random surface growth.
In its standard version, 
a nucleation of a layer with height one occurs with rate two per 
unit length. After a nucleation, the layer grows laterally in both 
directions from the nucleation point with unit speed. 
In addition, there occur nucleations with rate two on  
already existing layers. 
A discrete version of the PNG model was introduced in 
\cite{Jo2002} and studied further in \cite{Jo2002p}.
The discrete PNG model has a close relationship with the 
asymmetric simple exclusion process (ASEP) \cite{Jo2000,PS2002a}.

In \cite{BR2000,PS2000a,PS2000b},
the height fluctuation of the PNG model at a given time and 
a given position was studied. 
It turned out that the fluctuation strongly 
depends on the geometry of the models.
For instance, for the model in the infinite space, 
the fluctuation at the origin
is described by the GUE Tracy-Widom distribution 
in random matrix theory \cite{TW1994,Me1991}.
On the other hand, for the model in a half space with an 
external source at the origin, 
it is described by the GOE/GSE Tracy-Widom distributions 
\cite{TW1996} or by the Gaussian.
Models with other geometries are also considered in 
\cite{PS2000a,PS2000b}.
These are based on the results in 
\cite{BR2000,BR2001a,BR2001b,BR2001c}, 
in which the longest increasing subsequence problem 
with some symmetries are considered.

For the models in the infinite space,
multi-point equal time height fluctuations are studied 
by introducing the multi-layer version of the PNG model
in \cite{PS2002b} for the continuous time case and in 
\cite{Jo2002p} for the discrete time case. 
It is shown that the scaling limit is described by the 
Airy process, which is closely related to the dynamics of 
the rightmost particle of the Dyson's Brownian motion model
\cite{Dy1962}.

In this article we study the multi-point equal time height 
fluctuations of the PNG model in a half space with external 
source at the origin. As mentioned above, the
fluctuation at a single point has already been considered
in \cite{BR2001c,PS2000a}.
At the origin, it is given by the GOE/GSE Tracy-Widom distribution
whereas at other points, it is the GUE Tracy-Widom distribution. 
But the multi-point fluctuation has not been studied.
Following the ideas of \cite{Jo2001,PS2002b}, we introduce
the multi-layer PNG model.
In the scaling limit we will show that the crossover 
between the fluctuation at the origin and that in the bulk
corresponds to the orthogonal/symplectic to unitary transition at 
soft edge in random matrix theory \cite{FNH1999}. 

The outline of the paper is as follows.
In the next section, the model we consider is defined.
In section \ref{detkernel}, a determinantal process is considered.
Section \ref{scaling} deals with the scaling limit of the models. 
In section \ref{symplectic}, the results for the 
model without external source at the origin are summarized.
The standard PNG model in a half space can be considered by 
taking the limit for the discrete model. This is done in 
section \ref{continuous}. 
Some discussions, conjectures and Monte-Carlo results
are given in section \ref{discussions}, followed by the conclusion 
in the last section.

\setcounter{equation}{0}
\setcounter{theorem}{1}
\section{Model}
Let $r\in\N=\{0,1,2,\cdots\}$ and $t\in\N$ denote the discrete 
space and time coordinates respectively and $h(r,t)$ the 
height of the surface at position $r$ and at time $t$.
Our model lives in a half space $r\in\N$, but we can extend 
$h(r,t)$ to the whole space by setting $h(r,t) = h(-r,t)$ for 
$r\in\Z$, i.e. the height is symmetric under the reflection 
with respect to the origin.
We can further extend $h(r,t)$ to all $r\in\R$ by setting 
$h(r,t) = h([r],t)$. Then the height looks symmetric under the 
reflection with respect to $r=1/2$, 
i.e. $h(r,t) = h(1-r,t)$ for $r\in\R\setminus\Z$. 
The model considered in this article is described as follows.
Initially at time $t=0$ we have a flat substrate. 
At an odd time $t=1,3,\cdots$, nucleations occur at even sites
$r=0,\pm 2,\cdots,\pm t-1$.
At the origin, a nucleation of a height $k$ ($\in\N$) happens 
with the probability  $(1-\g\sqrt{q})(\g\sqrt{q})^k$ 
where $0<q<1$ and $0 \leq \g< 1/\sqrt{q}$.
For the height to be symmetric, nucleations  
at other points should be symmetric with respect to the origin; 
if the nucleation of a height $k$ occurs at $r$, so does at $-r$ as well.
Each independent nucleation with height $k$ occurs 
with probability $(1-q)q^k$.
At an even time $t=2,4,cdots$, nucleations occur at odd sites
$r=\pm 1,\pm 3, \cdots \pm t-1$. 
The nucleations are again synchronous with respect to the origin 
and a height of each independent nucleation is a 
geometric random variable with parameter $q$. 
In the meantime the steps grow laterally in both directions 
with unit speed. Sometimes a downstep and an upstep collide 
with each other, in which case the two steps merge to one with 
the higher height. See Fig. 1.
As is clear, the parameter $q$ is related to the frequency of 
nucleations in the bulk whereas 
the parameter $\g$ represents the strength of the external 
source at the origin. The bigger $\g$ is, the stronger 
the source is. In particular $\g=0$ corresponds to the model
without the external source at the origin.

Mathematically, our discrete PNG model can be defined by 
\begin{equation}
 h(r,t+1) = \text{max}(h(r-1,t),h(r,t),h(r+1,t)) + \o(r,t+1),
\label{PNGdef}
\end{equation}
with the initial condition $h(r,0)=0, r\in\Z$.
Here $\o$ is the random variable which takes a value in $\N$. 
$\o(r,t)=0$ if $t-r$ is even or if $|r|>t$, and
\begin{equation}
 w(i,j) = \o(i-j,i+j-1),
\end{equation}
$(i,j)\in\Z_+^2$ are geometric random variables.
In our model, all $w(i,j)$'s are not independent as in 
\cite{Jo2002p}, but a symmetry condition $w(i,j)=w(j,i)$ is
imposed. The parameters of the independent geometric random 
variables are given by
$q$ (resp. $\g\sqrt{q}$) for off-diagonal
(resp. diagonal) points, 
\begin{align}
 \P[w(i,j) = k] &= (1-q) q^k,             \quad 1 \leq j < i, \\
 \P[w(i,i) = k] &= (1-\g\sqrt{q}) (\g\sqrt{q})^k, \quad 1 \leq i,
\label{wii}
\end{align}
for $k\in\N$.
Introduction of $w(i,j)$ above is almost unnecessary for the 
following discussions, but useful to see the connection to 
the problems of last passage percolation and ASEP 
\cite{BR2000,Jo2000,BR2001c}.

In the following, we devote our most discussions to the 
two special choices of $\g$ where the detailed analysis of the 
model is possible.
One choice is $\g=1$, which is known to correspond to a 
critical point of the model \cite{BR2001c}. 
The other is $\g=0$, which corresponds to the model without 
the external source at the origin. 
In the following, the $\g=1$ (resp. $\g=0$) case will sometimes be 
referred to as the orthogonal (resp. symplectic) case because
it will turn out to be related to the orthogonal 
(resp. symplectic) to unitary transition of symmetries in random 
matrix theory. 
The analysis of the two cases are quite analogous, so that we 
mainly consider the $\g=1$ case in the sequel and summarize the 
results for the $\g=0$ case in section \ref{symplectic}.

We would like to know the statistical properties of $h(r,t)$.
As already mentioned in the introduction,
the fluctuation of the height at a single point is already known.
It is given by the GOE Tracy-Widom distribution
at the origin \cite{BR2001c} and by
the GUE Tracy-Widom distribution at other points \cite{PS2000a}.
In this article, we are interested in the crossover between these two.
Following \cite{Jo2002,PS2002b,Jo2002p}, 
we introduce the multi-layer version of the model which is 
defined as follows. In the original single-layer discrete PNG model, 
when two steps meet, the higher step takes over the lower one.
The information of the lower step is lost.
To keep this information, let us suppose that there was a second layer
below the original layer from the outset.
The height of the second layer remains $-1$ until a collision of 
steps happens at the first layer, at which time the nucleation 
for the second layer occurs just below the collision. 
The height of the nucleation is taken to be the height of the
overlapped region of the higher and the lower steps 
which collide at the first layer. See Fig. 1.
The steps in the second layer grows laterally in both directions 
with unit speed as in the first layer. The nucleations at the second 
layer are only due to collisions of steps at the first layer; 
there is no additional probabilistic nucleations. 
We denote the height of the second layer at position $r$ and 
at time $t$ as $h_1(r,t)$.
One can further suppose that there were infinitely many layers 
below the first layer, equally spaced at time 0
and successively construct $h_i(r,t)$ for $i=2,3,\cdots$. 
We also set $h(r,t)=h_0(r,t)$. This is the multi-layer PNG model.
An example of the whole height configurations is given in Fig. 2.

In the following, we mainly consider an even time, which 
is fixed to $t=M=2N$ until the asymptotics is considered.
There are infinitely many lines, but for time $t=M$ 
the number of lines which are not straight is at most $N$.
Moreover, there is a restriction, $h_i(M,M)=-i$($i=0,1,\cdots,N-1$).
For each jump with length $k$ of all layers, we associate a weight 
$(\sqrt{q})^k$. 
Since the whole configuration of the layers keep all information
of nucleations, the product of all weights of all jumps in the 
multi-layer PNG at $t=M$ is given by 
\begin{equation}
 \prod_{i+j\leq 2N+1,i>j} q^{w(i,j)} \prod_{i=1}^N (\sqrt{q})^{w(i,i)}.
\end{equation}

From now on, we change the way of looking at multi-layer heights.
The space coordinate $r$ in the original setting will be interpreted 
as the time coordinate.
The height coordinate will be interpreted as the space coordinate
and is represented as $x$.
We set $h_i(r,t) = x_{i+1}^r$ for $i=0,1,\cdots,N-1$ and
$r=0,1,\cdots,M$, and regard the height lines as paths of particles. 
Let us denote the collection of the positions of particles at 
time $r$ as $x^r=(x_1^r,\cdots,x_N^r)$ where $x_1^r > \cdots > x_N^r$,
and a collection of them for all times as $\bar{x} = (x^0,\cdots,x^M)$.

In what follows we use $\a=\sqrt{q}$.
Suppose that, when the the particle picture of heights are 
employed, the $x$ and $r$ axes are pointing the right and 
the up respectively.
Then at odd times, particles move to the right.
The transition weight of a particle from $x$ at time $2u$ 
to $y$ at time $2u+1$ is given by
\begin{align}
\label{phie}
 \p_{2u,2u+1}(x,y) 
 = 
 \begin{cases}
  (1-\a)\a^{y-x}, & y \geq x, \\
  0,                            & y < x.
 \end{cases}
\end{align}
At even times, they move to the left. The transition weight 
for each particle is
\begin{align}
\label{phio}
 \p_{2u-1,2u}(x,y) 
 =
 \begin{cases}
  0,                            & y > x ,\\
  (1-\a)\a^{x-y}, & y \leq x. 
 \end{cases}
\end{align}
Then, according to the Karlin-McGregor theorem \cite{KM1959a,KM1959b}, 
the weight of non-intersecting paths of $n$ particles which ends 
at $x_i^M=1-i$ is given by 
\begin{equation}
\label{GV}
 w_{n,M}(\bar{x}) 
 = 
 \prod_{r=0}^{M-1} \det (\p_{r,r+1}(x_i^r,x_j^{r+1}))_{i,j=1}^n,
\end{equation}
with the restriction $x_i^M=1-i$. 
The weight is written in a determinantal form, and hence 
the corresponding process is called a determinantal process.
Notice that the weight in (\ref{GV}) with $n=N$ is slightly 
different from the weight of the multi-layer PNG model because
some configurations contained in (\ref{GV}) 
are not allowed in the multi-layer PNG model where 
there are infinitely many particles.
On the other hand, if we take the limit $n\to\i$ from the beginning, 
there would appear another difficulty due to the infinity of the 
number of particles. Thus, for simplicity of the treatment, 
we approximate the weight of the multi-layer PNG model 
by (\ref{GV}) for the 
moment and take the limit $n\to\i$ afterwards when it becomes easy.
The simple determinantal expression in (\ref{GV}) allows us to 
perform a detailed analysis of the model.
If we took diagonal parameters to be more general as in (\ref{wii}),
the expression would become more complicated.

For later use, we introduce the generating function of $\p_{r,r+1}$,
\begin{align}
\label{fe}
 f_{2u}(z)   &= \sum_{k\in\Z}\p_{2u,2u+1}(k) z^k
              = \frac{1-\a}{1-\a z}, \\
\label{fo}
 f_{2u-1}(z) &= \sum_{k\in\Z}\p_{2u-1,2u}(k) z^k
             = \frac{1-\a}{1-\a/z}.
\end{align}
Let us denote the Fourier coefficient of a function $a(z)$ as 
$\hat{a}(k)$ like
\begin{equation}
 a(z) = \sum_{k\in\Z} \hat{a}(k) z^k.
\end{equation}
For instance one sees
\begin{align}
\label{pfe}
 \p_{2u,2u+1}(x,y) 
 &= 
 \hat{f}_{2u}(y-x), \\
\label{pfo}
 \p_{2u-1,2u}(x,y) 
 &=
 \hat{f}_{2u-1}(y-x) .
\end{align}
In addition, for a product,
\begin{equation}
 f_{r_1,r_2}(z) = \prod_{l=r_1}^{r_2} f_l(z),
\end{equation}
the Fourier coefficient, 
\begin{equation}
\label{Fc}
 \p_{r_1,r_2}(x,y)
 =
 \hat{f}_{r_1,r_2}(y-x),
\end{equation}
represents the transition weight of a particle 
between $(r_1,x)$ and $(r_2,y)$. Notice that when $r_1=r_2=r$ one has
\begin{equation}
 \p_{r,r}(x,y) = \delta_{xy}.
\end{equation}

Before going to the next section, a remark is in order.
Although we have employed the situation where the geometric 
random variables are given by a single parameter $q$, 
up to some point, it is easy to generalize our discussions to the case
where
\begin{align}
 \label{wa1}
 \P[w(i,j) = k] &= (1-a_i a_j) (a_i a_j)^k, \quad 1 \leq j < i, \\
 \label{wa2}
 \P[w(i,i) = k] &= (1-a_i) a_i^k, \quad 1 \leq i ,
\end{align}
with $0< a_i < 1$ for all $i$.
Then (\ref{fe}),(\ref{fo}) are replaced by
\begin{align}
\label{fea}
 f_{2u}(z)   &= \frac{1-a_{N+u+1}}{1-a_{N+u+1} z}, \\
\label{foa}
 f_{2u-1}(z) &= \frac{1-a_{N-u+1}}{1-a_{N-u+1}/z},
\end{align}
with the transition weight of a particle again given by
(\ref{pfe}),(\ref{pfo}).
It should be noted that our free parameters are only $a_j$'s;
for the case in \cite{Jo2002}, there are two sets of free 
parameters $a_j,b_j$.

\setcounter{equation}{0}
\setcounter{theorem}{1}
\section{Determinantal Process and Kernel}
\label{detkernel}
Let us consider the weight of non-intersecting paths given by
\begin{equation}
 w_{n,M}(\bar{x}) 
 = 
 \prod_{r=0}^{M-1} \det (\p_{r,r+1}(x_i^r,x_j^{r+1}))_{i,j=1}^n,
\end{equation}
where $x_i^M$ ($i=1,2,\cdots,n$) is fixed.
As for $x_i^0$, there is no restriction.
This is the main difference between our model and the 
model in \cite{Jo2002p} in which $x_i^0$ is also fixed.
The matrix element $\p_{r,r+1}(x_i^r,x_j^{r+1})$
is the transition weight of a particle between 
$(r,x_i^r)$ and $(r+1,x_j^{r+1})$.
Our main focus is on the special case in (\ref{phie}),(\ref{phio}), 
but the results of this section do not depend on a 
specific choice of $\p$.
For instance, when particles can hop only to nearest neighbor
sites for each time, the determinantal process is nothing but
the vicious walks studied in \cite{Ba2000,NF2002,NKT2003}. 
Hence our discussions below can immediately be
applied to the problems as well. 

The partition function is defined as
\begin{equation}
\label{par_fun}
 Z_{n,M} = \frac{1}{(n!)^M} \sum_{\bar{x}} w_{n,M}(\bar{x}).
\end{equation}
The probability of the non-intersecting paths reads
\begin{equation}
\label{pNM}
 p_{n,M}(\bar{x})
 =
 \frac{1}{(n!)^M Z_{n,M}} w_{n,M}(\bar{x}).
\end{equation}

Let $g(r,x)$ ($r=1,\cdots,M$) be some function.
In this section, we show 
\begin{proposition}
\begin{equation}
 \sum_{\bar{x}} \prod_{r=0}^{M-1}\prod_{j=1}^n (1+g(r,x_j^r)) p_{n,M}(\bar{x})
 =
 \sqrt{\det(1+K_1 g)},
\label{detK}
\end{equation} 
where the determinant on the right hand side is the Fredholm
determinant,
\begin{align}
 \det(1+K_1 g)
 &=
 \sum_{k=0}^{\i} \frac{1}{k!} 
 \sum_{r_1=1}^M \sum_{x_1} \sum_{j_1=1,2} \cdots 
 \sum_{r_k=1}^M \sum_{x_k} \sum_{j_k=1,2}
 g(r_1,x_1)\cdots g(r_k,x_k) \notag\\
 &\quad 
 \det(K_1(r_l,x_l;r_{l'},x_{l'})_{j_l,j_{l'}})_{l,l'=1}^k,
\end{align}
and $K_1$ is the $2\times 2$ matrix kernel,
\begin{equation}
\label{K1def}
 K_1(r_1,x_1;r_2,x_2)
 =
 \begin{bmatrix}
  S_1(r_1,x_1;r_2,x_2) & D_1(r_1,x_1;r_2,x_2) \\
  I_1(r_1,x_1;r_2,x_2) & S_1(r_2,x_2;r_1,x_1) 
 \end{bmatrix},
\end{equation}
with the matrix elements being
\begin{align}
\label{Sfin}
 S_1(r_1,x_1;r_2,x_2)
 &=
 \tilde{S}_1(r_1,x_1;r_2,x_2) - \phi_{r_1,r_2}(x_1,x_2), \\
\label{Stfin}
 \tilde{S}_1(r_1,x_1;r_2,x_2) 
 &=
 -\sum_{i,j=1}^n \p_{r_1,M}(x_1,x_i^M) (A_1^{-1})_{i,j} G_1(r_2,x_2;M,x_j^M),\\
\label{Ifin}
 I_1(r_1,x_1;r_2,x_2)
 &=
 \tilde{I}_1(r_1,x_1;r_2,x_2)  -G_1(r_1,x_1;r_2,x_2) ,\\ 
\label{Itfin}
 \tilde{I}_1(r_1,x_1;r_2,x_2)  
 &=
 -\sum_{i,j=1}^n G_1(r_1,x_1;M,x_i^M) (A_1^{-1})_{i,j} G_1(r_2,x_2;M,x_j^M) ,\\
\label{Dfin}
 D_1(r_1,x_1;r_2,x_2)
 &=
 \sum_{i,j=1}^n  \p_{r_1,M}(x_1,x_i^M) (A_1^{-1})_{i,j} \p_{r_2,M}(x_2,x_j^M).
\end{align}
Here
\begin{equation}
\label{Adef}
 (A_1)_{ij} 
 = 
 \sum_{y_1,y_2} \sgn (y_2-y_1) \p_{0,M}(y_1,x_i^M) \p_{0,M}(y_2,x_j^M),
\end{equation}
\begin{equation}
\label{G1def}
 G_1(r_1,x_1;r_2,x_2) 
 =
 \sum_{y_1,y_2} \sgn (y_2-y_1) \p_{0,r_1}(y_1,x_1) \p_{0,r_2}(y_2,x_2). 
\end{equation}
The expression for general $K_1(r_i,x_i;r_j,x_j)$
($i,j=1,2,\cdots,M$) has the same form as $K_1(r_1,x_1;r_2,x_2)$.
\end{proposition}

\vspace{3mm}\noindent
{\bf Remark.} 
The subscript 1 refers to the fact that this case is related to 
the orthogonal-unitary transition in random matrix theory.
If we take $g(r,x) = -\chi_{J_r}(x)$ ($r=1,2,\cdots, M$), 
where $\chi_{J_r}$ is the characteristic function of $J_r$, 
(\ref{detK}) gives the probability that there is no particle
on $J_1\times \cdots \times J_M$.
It should also be remarked that at this stage the infinite particles 
limit is taken easily. One only replaces the summation,
$\sum_{i,j=1}^n$, in (\ref{Stfin}),(\ref{Itfin}),(\ref{Dfin}) 
with $\sum_{i,j=1}^{\i}$.

\vspace{3mm}\noindent
{\bf Proof.} 
We derive (\ref{detK}) by
generalizing the methods of \cite{Jo2002p} and \cite{TW1998}. 
Let us define
\begin{equation}
 Z_{n,M}[g] 
 =
 \frac{1}{(n!)^M} \sum_{\bar{x}} \prod_{r=0}^{M-1}\prod_{j=1}^n 
 (1+g(r,x_j^r)) w_{n,M}(\bar{x}) .
\end{equation}
Remark $Z_{n,M}[0] = Z_{n,M}$ so that 
\begin{equation}
\label{ZZ}
 \sum_{\bar{x}} \prod_{r=0}^{M-1}\prod_{j=1}^n (1+g(r,x_j^r)) p_{n,M}(\bar{x})
 =
 \frac{Z_{n,M}[g]}{Z_{n,M}[0]}.
\end{equation}
By repeated use of the Heine identity,
\begin{equation}
\label{Heine}
 \frac{1}{n!}\sum_x \det(\phi_i(x_j))_{i,j=1}^n 
                    \det(\varphi_i(x_j))_{i,j=1}^n
 =
 \det\left(\sum_X \phi_i(X) \varphi_j(X)\right)_{i,j=1}^n ,
\end{equation}
one has
\begin{align}
 Z_{n,M}[g] 
 &=
 \sum_{x_1^0>\cdots > x_n^0}
 \det\Bigl( \sum_{X_1,\cdots,X_{M-1}} 
       (1+g(0,x_i^0))\phi_{0,1}(x_i^0,X_1)(1+g(1,X_1))\cdots  \notag\\
 &\quad 
       \phi_{M-2,M-1}(X_{M-2},X_{M-1})(1+g(M-1,X_{M-1})) 
       \phi_{M-1,M}(X_{M-1},x_j^M)
     \Bigr)_{i,j=1}^n .
\end{align}
Now, using another identity \cite{deB1955,TW1998},
\begin{equation}
 \left(\sum_{x_1>\cdots >x_n} \det(\phi_i(x_j))_{i,j=1}^n  \right)^2
 =
 \det\left( \sum_{y_1,y_2}\sgn (y_2-y_1) \phi_i(y_1)\phi_j(y_2)
     \right)_{i,j=1}^n ,
\end{equation}
one finds
\begin{align}
 Z_{n,M}[g]^2
 &=
 \det \Bigl(
   \sum_{y_1,y_2} \sgn(y_2-y_1) 
   \sum_{X_1,\cdots,X_{M-1}}(1+g(0,y_1))\phi_{0,1}(y_1,X_1)\cdots
                            \phi_{M-1,M}(X_{M-1},x_i^M) \notag\\
 &\quad 
   \sum_{\tilde{X}_1,\cdots,\tilde{X}_{M-1}}(1+g(0,y_2))
   \phi_{0,1}(y_2,\tilde{X}_1)\cdots
   \phi_{M-1,M}(\tilde{X}_{M-1},x_j^M) 
      \Bigr)_{i,j=1}^n  \notag\\
 &=
 \det \Bigl(
   \sum_{y_1,y_2} \sgn(y_2-y_1) \notag\\
 &\quad
   \sum_{X_0,\cdots,X_{M-1}}
     \phi_{0,0}(y_1,X_0)(1+g(0,X_0))\phi_{0,1}(X_0,X_1)\cdots
                            \phi_{M-1,M}(X_{M-1},x_i^M) \notag\\
 &\quad 
   \sum_{\tilde{X}_0,\cdots,\tilde{X}_{M-1}}
      \phi_{0,0}(y_2,\tilde{X}_0)(1+g(0,\tilde{X}_0))
      \phi_{0,1}(\tilde{X}_0,\tilde{X}_1)\cdots
      \phi_{M-1,M}(\tilde{X}_{M-1},x_j^M) 
       \Bigr)_{i,j=1}^n .
\end{align}
For the second equality we used $\phi_{0,0}(x,y)=\delta_{x,y}$.
We divide the $(i,j)$ element of the matrix in the above determinant 
into four parts with or without factors of the form
$g(r,X_r)$ or $g(\tilde{r},\tilde{X}_r)$.
We have
\begin{equation}
 Z_{n,M}[g]^2
 =
 \det \Bigl(
   (A_1)_{i,j} + (A_1^{(1)})_{i,j} + (A_1^{(2)})_{i,j} + (A_1^{(3)})_{i,j}
      \Bigr)_{i,j=1}^n,
\end{equation}
where $(A_1)_{i,j}$ is already defined in (\ref{Adef}) and
\begin{align}
 (A_1^{(1)})_{i,j}
 &=
 \sum_{y_1,y_2} \sgn (y_2-y_1)
 \sum_{l=1}^M \sum_{0\leq r_1 < \cdots < r_l < M}
 \sum_{X_1,\cdots,X_l} \phi_{0,r_1}(y_1,X_1)  \notag\\
 &\quad \prod_{s=1}^{l-1}
 g(r_s,X_s) \phi_{r_s,r_{s+1}}(X_s,X_{s+1}) 
 \cdot g(r_l,X_l) \phi_{r_l,M}(X_l,x_i^M)
 \cdot \phi_{0,M}(y_2,x_j^M) , 
\end{align}
\begin{align}
 (A_1^{(2)})_{i,j}
 &=
 \sum_{y_1,y_2} \sgn (y_2-y_1) ~
 \phi_{0,M}(y_1,x_i^M) 
 \sum_{l=1}^M \sum_{0\leq \tilde{r}_1 < \cdots < \tilde{r}_l < M}
 \sum_{\tilde{X}_1,\cdots,\tilde{X}_l} 
 \phi_{0,\tilde{r}_1}(y_2,\tilde{X}_1) \notag\\ 
 &\quad \prod_{s=1}^{l-1}
 g(\tilde{r}_s,\tilde{X}_s) \phi_{\tilde{r}_s,\tilde{r}_{s+1}}
 (\tilde{X}_s,\tilde{X}_{s+1}) 
 \cdot g(\tilde{r}_l,\tilde{X}_l) 
 \phi_{\tilde{r}_l,M}(\tilde{X}_l,x_j^M) , 
\end{align}
\begin{align}
 (A_1^{(3)})_{i,j}
 &=
 \sum_{y_1,y_2} \sgn (y_2-y_1)
 \sum_{l=1}^M \sum_{0\leq r_1 < \cdots < r_l < M}
 \sum_{X_1,\cdots,X_l} \phi_{0,r_1}(y_1,X_1) \notag\\
 &\quad \prod_{s=1}^{l-1}
 g(r_s,X_s) \phi_{r_s,r_{s+1}}(X_s,X_{s+1}) 
 \cdot g(r_l,X_l) \phi_{r_l,M}(X_l,x_i^M)
 \notag\\ 
 &\quad
 \sum_{m=1}^M 
 \sum_{0\leq \tilde{r}_1 < \cdots < \tilde{r}_m < M}
 \sum_{\tilde{X}_1,\cdots,\tilde{X}_{m}} 
 \phi_{0,\tilde{r}_1}(y_2,\tilde{X}_1) \notag\\
 &\quad \prod_{s=1}^{m-1}
 g(\tilde{r}_s,\tilde{X}_s) \phi_{\tilde{r}_s,\tilde{r}_{s+1}}
 (\tilde{X}_s,\tilde{X}_{s+1}) 
 \cdot g(\tilde{r}_m,\tilde{X}_m) 
 \phi_{\tilde{r}_m,M}(\tilde{X}_m,x_j^M). 
\end{align}

Let us define
\begin{align}
 \vp(r_1,x_1;r_2,x_2) &= g(r_1,x_1) \phi_{r_1,r_2}(x_1,x_2), \\
 \tilde{\vp}(r_1,x_1;r_2,x_2) &=  \phi_{r_1,r_2}(x_1,x_2) g(r_2,x_2),
\end{align}
and 
\begin{align}
\label{defphii}
 \phi_i(r,x)
 &=
 \phi_{r,M}(x,x_i^M), \\
 \label{defGi}
 (G_1)_i(r,x) &= G_1(r,x;M,x_i^M), \\
 \label{defpsi}
 \psi_i(r,x)
 &=
 \sum_{l=1}^M \sum_{r_2,\cdots, r_l}
 \sum_{X_2,\cdots,X_l} \phi_{r,r_2}(x,X_2) 
 \prod_{s=2}^{l-1}
 \vp(r_s,X_s;r_{s+1},X_{s+1})
 \cdot \vp(r_l,X_l;M,x_i^M).
\end{align} 
Then $(A_1^{(1)})_{i,j}$ can be rewritten as 
\begin{align}
 (A_1^{(1)})_{i,j}
 &=
 \sum_{l=1}^M \sum_{r_1=0}^{M-1} \cdots \sum_{r_l=1}^{M-1}
 \sum_{X_1,\cdots,X_l} 
 G_1(r_1,X_1;M,x_j^M) 
 g(r_1,X_1)\cdot \phi_{r_1,r_2}(X_1,X_2) \notag\\
 &\quad
 \prod_{s=2}^{l-1} \vp(r_s,X_s;r_{s+1},X_{s+1}) 
 \cdot \vp(r_l,X_l;M,x_i^M) \notag\\
 &=
 \sum_{r,x} g(r,x) \psi_i(r,x) G_1(r,x;M,x_j^M) \notag\\ 
 &=
 \sum_{r,x} g\psi_i \cdot (G_1)_j  .
\label{A11}
\end{align}
In the last expression, the dependence on $r,x$ is omitted
for notational simplicity.
Noticing the antisymmetry of $G_1$,
\begin{equation}
 G_1(r_1,x_1;r_2,x_2) = -G_1(r_2,x_2;r_1,x_1),
\end{equation}
we can also rewrite $(A_1)_{i,j}^{(2)}$ and $(A_1)_{i,j}^{(3)}$ as 
\begin{align}
\label{A12}
 (A_1^{(2)})_{i,j}
 &=
 -\sum_{r,x} g\psi_j \cdot (G_1)_i ,\\ 
 \label{A13}
 (A_1^{(3)})_{i,j}
 &=
 -\sum_{r,x} g\psi_j \cdot G_1(g\psi_i),
\end{align}
where we used the notation,
\begin{equation}
 (G_1 f)(r,x) = \sum_{r_1,x_1} G_1(r,x;r_1,x_1) f(r_1,x_1).
\end{equation}
Hence we find
\begin{equation}
 Z_{n,M}[g]^2
 =
 \det\left((A_1)_{i,j}+ \sum_{r,x} 
 (  g\psi_i\cdot (G_1)_j -g\psi_j \cdot (G_1)_i 
  - g \psi_j\cdot G_1(g\psi_i))\right)_{i,j=1}^n. 
\end{equation}
In view of (\ref{ZZ}), we divide this by 
\begin{equation}
Z_{n,M}[0]^2 = \det A_1.
\end{equation}
In the determinant this corresponds to multiplying by $A_1^{-1}$,
say from the left. Let $\eta_i = \sum_j (A_1^{-1})_{i,j} \psi_j,
     \Psi_i = \sum_j (A_1^{-1})_{i,j} (G_1)_j$. Then 
\begin{equation}
 \left(\frac{Z_{n,M}[g]}{Z_{n,M}[0]}\right)^2
 =
 \det\left( \delta_{i,j}+ \sum_{r,x} 
 (g\eta_i\cdot (G_1)_j -g\psi_j \cdot \Psi_i - g \psi_j\cdot
 G_1(g\eta_i))\right)_{i,j=1}^n.
\end{equation}
One notices that, if one introduces
\begin{equation}
 B(i;r,x) = (-g \Psi_i -g G_1(g\eta_i), g\eta_i) , \quad
  C(r,x;i) = \binom{\psi_i}{(G_1)_i},
\end{equation}
this is equal to
\begin{equation}
 \det\left(\delta_{i,j}+\sum_{r,x}B(i;r,x) C(r,x;j) \right)_{i,j=1}^n
 =
 \det(1+BC).
\end{equation}
Now we use a simple fact
\begin{equation}
 \det(1+BC) = \det(1+CB).
\end{equation}
The determinant on the left hand side is the determinant of a
matrix of finite rank, whereas the one on the right hand side 
is the Fredholm determinant.
We see
\begin{equation}
 \det(1+CB) \notag\\
 =
 \det
  \begin{bmatrix}
   1-\sum_i \psi_i \otimes g \Psi_i -\sum_i \psi_i \otimes gG_1(g\eta_i)
   & 
   \sum_i \psi_i \otimes g\eta_i \\
   -\sum_i (G_1)_i \otimes g \Psi_i -\sum_i (G_1)_i \otimes gG_1(g\eta_i) 
   &
   1 + \sum_i (G_1)_i \otimes g\eta_i   
  \end{bmatrix},
\end{equation}
where we used the notation $a\otimes b$ for 
the operator with kernel of the form $a(r_1,x_1) b(r_2,x_2)$.
The matrix in the determinant may be rewritten as a product 
of two matrices, leading to
\begin{equation}
 \det\left(
  \begin{bmatrix}
   1-\sum_i \psi_i \otimes g \Psi_i & \sum_i \psi_i \otimes g\eta_i \\
   -\sum_i (G_1)_i \otimes g \Psi_i 
   - G_1g & 1 + \sum_i (G_1)_i \otimes g\eta_i   
  \end{bmatrix}  
  \begin{bmatrix}
   1     & 0 \\
   G_1 g & 1 
  \end{bmatrix}  
 \right) .
\end{equation}
The determinant of the right matrix is one, so that this equals
\begin{equation}
 \det\left(1+
  \begin{bmatrix}
   -\sum_i \psi_i \otimes g \Psi_i & \sum_i \psi_i \otimes g\eta_i \\
   -\sum_i (G_1)_i \otimes g \Psi_i - G_1 g & \sum_i (G_1)_i \otimes g\eta_i   
  \end{bmatrix}  
 \right) .
\end{equation}
Let us remember the definition of $\psi_i$ in (\ref{defpsi})
and notice
\begin{align}
 \psi_i  &= \sum_{l=1}^M \tilde{\vp}^{*(l-1)} * \phi_i
          = (1-\tilde{\vp})^{-1} * \phi_i , \\
 g\psi_i &= \sum_{l=1}^M \vp^{*(l-1)} * (g \phi_i)
          = (1-\vp)^{-1} * (g \phi_i) ,
\end{align}
where
\begin{equation}
 (f_1 * f_2)(r_1,x_1;r_2,x_2)
 =
 \sum_{r,x}f_1(r_1,x_1;r,x) f_2(r,x;r_2,x_2).
\end{equation}
The determinant we are considering is now
\begin{equation}
 \det\left(1+
  \begin{bmatrix}
   (1-\tilde{\vp})^{-1} & 0 \\
   0 & 1 
  \end{bmatrix}  
  \begin{bmatrix}
   -\sum_i \phi_i \otimes g \Psi_i & \sum_i \phi_i 
   \otimes g (A_1^{-1}\phi)_i \\
   -\sum_i (G_1)_i \otimes g \Psi_i 
   - G_1g & \sum_i (G_1)_i \otimes g (A_1^{-1}\phi)_i   
  \end{bmatrix}  
  \begin{bmatrix}
   1 & 0 \\
   0 & (1-\,^t \vp)^{-1}  
  \end{bmatrix}  
 \right) .
\end{equation}
Multiplying from the left by
\begin{equation}
 1=\det  
 \begin{bmatrix}
  1-\tilde{\vp} & 0 \\
  0 & 1 
 \end{bmatrix}  
\end{equation}
and from the right by
\begin{equation}
 1=\det  
 \begin{bmatrix}
  1 & 0 \\
  0 & 1-\,^t\vp   
 \end{bmatrix}, 
\end{equation}
one gets
\begin{align}
 &\quad
 \left(\frac{Z_{n,M}[g]}{Z_{n,M}[0]}\right)^2 \notag\\
 &=
 \det\Biggl(
  \begin{bmatrix}
   1-\tilde{\vp} & 0 \\
   0 & 1- \,^t\vp 
  \end{bmatrix}
 \notag\\
 &\quad
  +
  \begin{bmatrix}
   -\sum_{i,j} \phi_i \otimes  (A_1^{-1})_{i,j} (G_1)_j 
   & \sum_{i,j} \phi_i \otimes (A_1^{-1})_{i,j} \phi_j \\
   -\sum_{i,j} (G_1)_i \otimes (A_1^{-1})_{i,j} (G_1)_j  - G_1 
   & \sum_{i,j} (G_1)_i \otimes (A_1^{-1})_{i,j} \phi_j
  \end{bmatrix} g 
 \Biggr) \notag\\
 &=
 \det\left(1+
  \begin{bmatrix}
   -\sum_{i,j} \phi_i \otimes (A_1^{-1})_{i,j} (G_1)_j -\phi  
   & \sum_{i,j} \phi_i \otimes (A_1^{-1})_{i,j} \phi_j \\
   -\sum_{i,j} (G_1)_i \otimes (A_1^{-1})_{i,j} (G_1)_j  - G_1 
   & \sum_{i,j} (G_1)_i \otimes (A_1^{-1})_{i,j}\phi_j -\,^t \phi
  \end{bmatrix} g 
 \right) . 
\end{align}
Recalling the definitions of $\phi_i$ and $(G_1)_i$
in (\ref{defphii}) and (\ref{defGi}), we see that (\ref{detK}) holds.
\qed

\setcounter{equation}{0}
\setcounter{theorem}{1}
\section{Scaling Limit}
\label{scaling}
In this section we consider the scaling limit, where the 
universal properties of the model are expected to appear.

\subsection{Generating Functions}
\label{gen}
To study the asymptotics of the kernel, we  
compute the generating functions of $\tilde{S}_1,\tilde{I}_1,D_1$.

First let us recall some basic facts 
about a Toeplitz matrix \cite{MW1973}.
A Toeplitz matrix is a matrix $A$ of the form,
\begin{equation}
\label{matA}
 A 
 =
 \begin{bmatrix}
  a_0    & a_{-1} & a_{-2} & \cdots \\
  a_1    & a_0    & a_{-1} &        \\
  a_2    & a_1    & a_0    & \ddots \\
  \vdots &        & \ddots & \ddots 
 \end{bmatrix}.
\end{equation}
It is useful to define a function,
\begin{equation}
 a(z) = \sum_{k\in\Z} a_k z^k,
\end{equation}
which is called the symbol of the Toeplitz matrix $A$.
Conversely, for a given function $a(z)$, set 
$a_k=\hat{a}(k)$, the Fourier coefficient of the symbol $a(z)$.
Then one can define the corresponding Toeplitz matrix (\ref{matA}), 
which will be denoted by $T(a)$.
The inverse of $A$ can be given in a compact fashion,  
if the symbol $a(z)$ has winding number zero and is 
Wiener-Hopf factorizable as
\begin{equation}
\label{afac}
 a(z) = a_+(z) a_-(z)
\end{equation}
with
\begin{align}
 a_+(z) &= \sum_{k=0}^{\i} \hat{a}_+(k) z^k, \\
 a_-(z) &= \sum_{k=0}^{\i} \hat{a}_-(k) z^{-k}.
\end{align}
In fact for a factorization (\ref{afac}) one has
\begin{equation}
 A = T(a) = T(a_-) T(a_+).
\end{equation}
In addition, the inverse of the matrix $T(a_{\pm})$ is 
simply given by
\begin{equation}
 T(a_{\pm})^{-1} = T\left(\frac{1}{a_{\pm}}\right).
\end{equation}
Hence the inverse of the matrix $A$ is
\begin{equation}
\label{invA}
 A^{-1} = T \left(\frac{1}{a_+}\right) T\left(\frac{1}{a_-}\right).
\end{equation}

To obtain the generating functions of $\tilde{S}_1,\tilde{I}_1,D_1$, 
we need the inverse of the matrix $A_1$ (\ref{Adef}).
As remarked below (\ref{G1def}), for the multi-layer PNG model, 
the matrix $A_1$ should be taken to be an infinite dimensional one, 
which turns out to be a Toeplitz matrix. If we set
\begin{equation}
 (A_1)_{ij} = \hat{a}(i-j),
\end{equation}
the symbol is computed as
\begin{align}
\label{syma1}
 a_1(z)
 &=
 \sum_{k\in\Z} \hat{a}(k) z^k \notag\\
 &=
 \sum_{k,y_1,y_2}
 \hat{f}_{0,M}(1-k-y_1) z^{-1+k+y_1} 
 \cdot \sgn(y_2-y_1) z^{y_2-y_1} 
 \cdot \hat{f}_{0,M}(1-y_2) z^{1-y_2} \notag\\
 &=
 f_{0,M}\left(\frac{1}{z}\right) s_1(z) f_{0,M}(z),
\end{align}
where 
\begin{equation}
 s_1(z) = \sum_{k\in\Z} \sgn (k) z^k = \frac{1+z}{1-z}.
\end{equation}
In the second equality of (\ref{syma1}), we use (\ref{Fc})
and (\ref{Adef}).
The winding number of $a_1(z)$ is not zero and hence the 
formula (\ref{invA}) does not hold as it is in this case.
But the difficulty can be overcome fairly easily. 
Let us first notice that $s_1(z)$ can be written in two ways as
\begin{equation}
 s_1(z) = s_{1+}^{(1)}(z) s_{1-}^{(1)}(z)  
        = s_{1+}^{(2)}(z) s_{1-}^{(2)}(z) ,
\end{equation}
with
\begin{align}
 s_{1+}^{(1)}(z) &= \frac{1+z}{1-z},  \quad s_{1-}^{(1)}(z) = 1, \\
 s_{1+}^{(2)}(z) &= 1,   \qquad\quad  s_{1-}^{(2)}(z) = -\frac{1+1/z}{1-1/z}.
\end{align}
Correspondingly, if we define
\begin{align}
 a_{1+}^{(i)}(z) = f_{0,M;-}\left(\frac{1}{z}\right) 
                   f_{0,M;+}(z) s_{1+}^{(i)}(z), \\
 a_{1-}^{(i)}(z) = f_{0,M;+}\left(\frac{1}{z}\right) 
                   f_{0,M;-}(z) s_{1-}^{(i)}(z),
\end{align}
for $i=1,2$, we have
\begin{equation}
 a_1(z) =  a_{1+}^{(1)}(z) a_{1-}^{(1)}(z) = a_{1+}^{(2)}(z) a_{1-}^{(2)}(z).
\end{equation}
Then it is confirmed that the inverse $A_1^{-1}$ is given by
\begin{equation}
 A_1^{-1}
 =
 \frac12\left\{ T\left(\frac{1}{a_{1+}^{(1)}}\right) 
                T\left(\frac{1}{a_{1-}^{(1)}}\right)
               +T\left(\frac{1}{a_{1+}^{(2)}}\right) 
                T\left(\frac{1}{a_{1-}^{(2)}}\right) 
        \right\},
\end{equation}
which is the substitute for (\ref{invA}).

Now one can compute the generating function,
\begin{align}
 &\quad
 \sum_{i,j=1}^{\i} z_1^{1-i} (A_1^{-1})_{i,j} z_2^{j-1} \notag\\
 &=
 \sum_{k=1}^{\i} \frac12
 \Biggl\{ 
  \sum_{i\in\Z} z_1^{-i+k} \widehat{\left(1/a_{1+}^{(1)}\right)}(i-k) ~
  \sum_{j\in\Z} \widehat{\left(1/a_{1-}^{(1)}\right)}(k-j) ~z_2^{j-k} 
 \notag\\ 
 &\quad  +
  \sum_{i\in\Z} z_1^{-i+k} \widehat{\left(1/a_{1+}^{(2)}\right)}(i-k)~
  \sum_{j\in\Z} \widehat{\left(1/a_{1-}^{(2)}\right)}(k-j) ~ z_2^{j-k} 
 \Biggr\} \left(\frac{z_2}{z_1}\right)^{k-1} \notag\\
 &=
 \frac{1}{1-z_2/z_1}\frac12
 \left\{ \frac{1}{a_{1+}^{(1)}(\fz) a_{1-}^{(1)}(\fw)}
         +
         \frac{1}{a_{1+}^{(2)}(\fz) a_{1-}^{(2)}(\fw)}\right\} \notag\\
 &=
 \frac{z_1}{z_1-z_2}
 \frac{1}{f_{0,M;-}(z_1)f_{0,M;+}(\fz)f_{0,M;+}(z_2)f_{0,M;-}(\fw)} 
 \notag\\
 &\quad \times
 \frac12 \left\{ \frac{1}{s_{1+}^{(1)}(\fz)s_{1-}^{(1)}(\fw)}+
                 \frac{1}{s_{1+}^{(2)}(\fz)s_{1-}^{(2)}(\fw)} \right\},
\label{zAz}
\end{align}
which is valid for $|z_1|>|z_2|$. 
In the first equality we use a fact that 
$\widehat{\left(1/a_{1+}^{(i)}\right)}(k)=0$ for $k<0$ and 
$\widehat{\left(1/a_{1-}^{(i)}\right)}(k)=0$ for $k>0$ ($i=1,2$). 

We also introduce the generating function of the function $G_1$ 
in (\ref{G1def}). Note that as a function of $x_1,x_2$, $G_1$ only 
depends on the difference $x_1-x_2$, so that 
we can set
\begin{equation}
 G_1(r_1,x_1;r_2,x_2) = (\hat{G}_1)_{r_1,r_2}(x_1-x_2).
\end{equation}
Then one has
\begin{equation}
 (G_1)_{r_1,r_2}(z)
 =
 \sum_{k\in\Z} (\hat{G}_1)_{r_1,r_2}(k) z^k 
 =
 f_{0,r_1}(z) s_1\left(\frac{1}{z}\right) 
 f_{0,r_2}\left(\frac{1}{z}\right)
 =
 -f_{0,r_1}(z) s_1(z)  f_{0,r_2}\left(\frac{1}{z}\right)
\end{equation}

With these preparations, 
calculations of the generating functions of $\tilde{S}_1,\tilde{I}_1,D_1$
are not difficult. We have
\begin{align}
 &\quad
 \tilde{\Sg}_1(r_1,z_1;r_2,z_2) 
 =
 \sum_{x_1,x_2\in\Z} \tilde{S}_1(r_1,x_1;r_2,x_2) z_1^{x_1} z_2^{-x_2} \notag\\
 &=
 -\sum_{x_1,x_2} \sum_{i,j} \hat{f}_{r_1,M}(1-i-x_1) z_1^{-1+i+x_1} 
 \cdot  z_1^{1-i} (A_1^{-1})_{i,j} z_2^{j-1} 
 \cdot (\hat{G}_1)_{r_2,M}\left(x_2-1+j\right)z_2^{-x_2+1-j} \notag\\
 &=
 -f_{r_1,M}\left(\fz\right)\cdot \sum_{i,j} z_1^{1-i} 
 (A_1^{-1})_{i,j} z_2^{j-1} 
 \cdot (G_1)_{r_2,M}\left(\fw\right) \notag\\
 &= 
 \frac{z_1}{z_1-z_2} \frac{f_{r_1,M;-}(\fz) f_{0,r_2;+}(\fw) f_{0,M;-}(z_2)}
                          {f_{0,M;-}(z_1) f_{0,r_1;+}(\fz) f_{r_2,M;-}(\fw)}
 \frac12\left\{ \frac{s_{1+}^{(1)}(\fw)}{s_{1+}^{(1)}(\fz)}
               +\frac{s_{1+}^{(2)}(\fw)}{s_{1+}^{(2)}(\fz)}\right\}, \notag\\
 &=
\label{Sgg1}
 \frac{(1-\a)^{2(u_2-u_1)} (1-\a/z_1)^{N+u_1} (1-\a z_2)^{N-u_2}}
      {(1-\a z_1)^{N-u_1} (1-\a/z_2)^{N+u_2}} 
 \frac{z_1}{z_1-z_2} 
 \left\{1 + \frac{z_1-z_2}{(1+z_1)(z_2-1)} \right\}, 
\end{align}
\begin{align}
 &\quad
 \tilde{\Ig}_1(r_1,z_1;r_2,z_2) 
 = 
 \sum_{x_1,x_2\in\Z} \tilde{I}_1(r_1,x_1;r_2,x_2) z_1^{x_1} z_2^{-x_2} \notag\\
 &=
 -(G_1)_{r_1,M}(z_1)  \cdot \sum_{i,j} z_1^{1-i} (A_1^{-1})_{i,j} z_2^{j-1} 
 \cdot (G_1)_{r_2,M}\left(\fw\right) \notag\\
 &=
 \frac{z_1}{z_1-z_2} \frac{f_{0,r_1;+}(z_1) f_{0,M;-}(\fz) 
                           f_{0,r_2;+}(\fw) f_{0,M;-}(z_2)}
                          {f_{r_1,M;-}(z_1) f_{r_2,M;-}(\fw)}
 \notag\\
 &\quad \times
 \frac12\left\{ s_{1-}^{(1)}\left(\fz\right)s_{1+}^{(1)}\left(\fw\right)
               +s_{1-}^{(2)}\left(\fz\right)s_{1+}^{(2)}\left(\fw\right)
        \right\}, 
 \notag\\
 &=
\label{Igg1}
 \frac{(1-\a)^{2(u_1+u_2)} (1-\a/z_1)^{N-u_1} (1-\a z_2)^{N-u_2}}
      {(1-\a z_1)^{N+u_1} (1-\a/z_2)^{N+u_2}} 
 \frac{z_1}{z_1-z_2}
 \left\{ \frac12\frac{z_2+1}{z_2-1}-\frac12\frac{1+z_1}{1-z_1} \right\}, 
\end{align}
\begin{align}
 &\quad
 \Dg_1(r_1,z_1;r_2,z_2) 
 = 
 \sum_{x_1,x_2\in\Z} D(r_1,x_1;r_2,x_2) z_1^{x_1} z_2^{-x_2} \notag\\
 &=
 -f_{r_1,M}\left(\fz\right)  
 \cdot \sum_{i,j} z_1^{1-i} (A_1^{-1})_{i,j} z_2^{j-1} 
 \cdot f_{r_2,M}(z_2) \notag\\
 &=
 \frac{z_1}{z_1-z_2} \frac{f_{r_1,M;-}(\fz) f_{r_2,M;-}(z_2)}
                          {f_{0,M;-}(z_1)   f_{0,r_1;+}(\fz) 
                           f_{0,r_2;+}(z_2) f_{r_2,M;-}(\fw)}
 \frac12\left\{ \frac{1}{s_{1+}^{(1)}(\fz)s_{1-}^{(1)}(\fw)}
               +\frac{1}{s_{1+}^{(2)}(\fz)s_{1-}^{(2)}(\fw)}\right\} \notag\\
 &=
\label{Dgg1}
 \frac{(1-\a/z_1)^{N+u_1} (1-\a z_2)^{N+u_2}}
      {(1-\a)^{2(u_1+u_2)} (1-\a z_1)^{N-u_1} (1-\a/z_2)^{N-u_2}} 
 \frac{z_1}{z_1-z_2}
 \left\{ \frac12\frac{z_1-1}{z_1+1}-\frac12\frac{1-z_2}{1+z_2} \right\}.
\end{align}
In the last expressions of these, we have set $r_i=2u_i$ for $i=1,2$.

\subsection{Bulk}
\label{bulk}
We are now in a position to study the asymptotic behaviors of the model.
The thermodynamic limit shape is already known to be 
\begin{equation}
\label{shape}
 h(r=2\b_0 N,t=2N) /N 
 \sim 
 \frac{2\a}{1-\a^2} \left( \a+\sqrt{1-\b_0^2} \right),
\end{equation}
where $0<\b_0<1$ is fixed \cite{Ro1981}.
As for the fluctuation around this limit shape, 
the KPZ theory tells us that the correlation
survives for $O(N^{1/3})$ along the height direction and 
$O(N^{2/3})$ along the $r$ direction.

Let us define the scaled height variable as
\begin{equation}
\label{scaledH}
 H_N(\t,\b_0) 
 = 
 \frac{h(r=2\b_0 N+2c(\b_0)N^{\frac23}\t,t=2N)
       -a(\b_0+\frac{c(\b_0)\t}{N^{1/3}}) N}
      {d(\b_0) N^{\frac13}} ,
\end{equation}
where
\begin{align}
\label{defa}
 a(\b) &= \frac{2\a}{1-\a^2} \left( \a+\sqrt{1-\b^2} \right),\\
 d(\b) &= \frac{\a^{\frac13}}{(1-\a^2)(1-\b^2)^{\frac16}}
             (\sqrt{1+\b}+\a\sqrt{1-\b})^{\frac23}
             (\sqrt{1-\b}+\a\sqrt{1+\b})^{\frac23}, \\
 c(\b) &= \a^{-\frac13} (1-\b^2)^{\frac23}
             (\sqrt{1+\b}+\a\sqrt{1-\b})^{\frac13}
             (\sqrt{1-\b}+\a\sqrt{1+\b})^{\frac13}.
\end{align}
Remark that $a(\b)$ just comes from the limit shape (\ref{shape}) 
whereas $d(\b)$ and $c(\b)$ are taken so that the results
become simple.
In this subsection, we show
that the fluctuation of the model in the bulk 
is described by the Airy process \cite{PS2002b,Jo2002p}. 
Namely, we show

\begin{proposition}
\begin{equation}
\label{detKlim2}
 \lim_{N\to\i} 
 \P[H_N(\t_1,\b_0) \leq s_1, \cdots , H_N(\t_m,\b_0) \leq s_m]
 =
 \det(1+\K_2 \G).
\end{equation}
The determinant on the right hand side is the Fredholm determinant,
\begin{align}
 \det(1+\K_2 \G) 
 &=
 \sum_{k=0}^{\i} \frac{1}{k!} 
 \sum_{\t_1=1}^m \int d\xi_1 \cdots \sum_{\t_k=1}^m \int d\xi_k 
 \G(\t_1,\xi_1) \cdots \G(\t_k,\xi_k)  
 \notag\\
 &\quad 
 \det(\K_2(\t_l,\xi_l;\t_{l'},\xi_{l'}))_{l,l'=1}^k,
\end{align}
where $\G(\t_j,\xi)=-\chi_{(s_j,\i)}(x)$ ($j=1,2,\cdots, m$) 
and $\K_2$ is a scalar kernel,
for which a representative element is given by 
\begin{equation}
\label{extAi}
 \K_2(\t_1,\xi_1;\t_2,\xi_2)
 =
 \tilde{\K}_2(\t_1,\xi_1;\t_2,\xi_2)-\Phi_{\t_1,\t_2}(\xi_1,\xi_2), 
\end{equation}
with 
\begin{align}
 \label{K2def}
 \tilde{\K}_2(\t_1,\xi_1;\t_2,\xi_2)
 &=
 \int_0^{\i} d\l e^{-\l(\t_1-\t_2)} \Ai(\xi_1+\l) \Ai(\xi_2+\l) ,\\
 \label{Phidef}
 \Phi_{\t_1,\t_2}(\xi_1,\xi_2)
 &=
 \begin{cases}
  0, & \t_1 \geq \t_2 ,\\
  \int_{-\i}^{\i} d\l e^{-\l(\t_1-\t_2)} \Ai(\xi_1+\l) \Ai(\xi_2+\l), 
  & \t_1 < \t_2 .
\end{cases}
\end{align}
\end{proposition}

\vspace{3mm}\noindent
{\bf Remark.} 
The kernel (\ref{extAi}) is called the extended Airy kernel and 
had already appeared in the context of Dyson's
Brownian motion model \cite{Mac1994,NF1998}.
We also remark that some properties of the Airy process have been discussed 
in \cite{PS2002b,Jo2002p,TW2003p,AvM2003p}.
For the special case of a single time, we have
\begin{equation}
\label{HF2}
 \lim_{N\to\i} \P[H_N(\t,\b_0) \leq \xi ]
 =
 F_2(\xi),
\end{equation}
where $F_2(\xi)$ is the GUE Tracy-Widom distribution \cite{TW1994}.
In the language of the last passage percolation,
(\ref{HF2}) means that the fluctuation of the last passage time 
at an off-diagonal point is described by the GUE Tracy-Widom 
distribution. For the nonsymmetric case, the statement 
has been already established in \cite{Jo2000}. But for our 
symmetric case, this has not seem to be shown although it 
was already expected based on a physical argument in \cite{PS2000a}.
Remember also that the fluctuation at an off-diagonal point 
has not been handled by the techniques using the 
Riemann-Hilbert method \cite{BR2001c}.

\vspace{3mm}\noindent
{\bf Proof.} 
Let us derive (\ref{detKlim2}).
We first remark that (\ref{K2def}) has a double contour
integral representation,
\begin{equation}
 \tilde{\K}_2(\t_1,\xi_1;\t_2,\xi_2) \notag\\
 =
 -\frac{1}{4\pi^2}\int_{\text{Im}w_1=\eta_1} dw_1 
                  \int_{\text{Im}w_2=\eta_2} dw_2 
 \frac{ e^{i\xi_1 w_1 + i\xi_2 w_2 + \frac{i}{3}(w_1^3+w_2^3)}}
 {\tau_2-\tau_1+i(w_1+w_2)},
\end{equation}
when $\eta_1+\eta_2+\t_1-\t_2>0,\eta_1,\eta_2>0$, 
and that (\ref{Phidef}) can be computed as
\begin{equation}
 \Phi_{\t_1,\t_2}(\xi_1,\xi_2)
 =
 \frac{1}{\sqrt{4\pi (\t_2-\t_1)}}
 e^{-\frac{(\xi_2-\xi_1)^2}{4(\t_2-\t_1)}
    -\frac12(\t_2-\t_1)(\xi_1+\xi_2) 
    +\frac{1}{12}(\t_2-\t_1)^3}
\end{equation}
when $\t_2 > \t_1$. These are already given in \cite{Jo2002p}.

Using the generating functions in the previous subsection, 
we get the double contour integral formula for $\tilde{S}_1,\tilde{I}_1,D_1$,
\begin{align}
 \label{Sint}
 \tilde{S}_1(r_1,x_1;r_2,x_2) 
 &=
 \frac{1}{(2\pi i)^2}
 \int_{C_{R_1}} \frac{dz_1}{z_1^{x_1+1}} \int_{C_{R_2}} dz_2 z_2^{x_2-1} 
 \tilde{\Sg}_1(r_1,z_1;r_2,z_2), \\
 \label{Iint}
 \tilde{I}_1(r_1,x_1;r_2,x_2) 
 &=
 \frac{1}{(2\pi i)^2}
 \int_{C_{R_1}} \frac{dz_1}{z_1^{x_1+1}} \int_{C_{R_2}} dz_2 z_2^{x_2-1} 
 \tilde{\Ig}_1(r_1,z_1;r_2,z_2), \\
 \label{Dint}
 D_1(r_1,x_1;r_2,x_2) 
 &=
 \frac{1}{(2\pi i)^2}
 \int_{C_{R_1}} \frac{dz_1}{z_1^{x_1+1}} 
 \int_{C_{R_2}} dz_2 z_2^{x_2-1} \Dg_1(r_1,z_1;r_2,z_2) .
\end{align}
Here $C_R$ denotes a contour enclosing the origin anticlockwise
with radius $R$. We have taken $R_1 > R_2$ to agree with the computation in  
(\ref{zAz}). For $\tilde{I}_1$, exchanging the radius of the contours
of $z_1,z_2$ gives
\begin{align}
 &\quad 
 \tilde{I}_1(r_1,x_1;r_2,x_2) \notag\\
 &= 
 \frac{1}{(2\pi i)^2}
 \int_{C_{R_2}} \frac{dz_1}{z_1^{x_1+1}} 
 \int_{C_{R_1}} dz_2 z_2^{x_2-1} 
 \tilde{\Ig}_1(r_1,z_1;r_2,z_2) -\frac{1}{2\pi i}
 \int_{C_{R_2}} dz_2 z_2^{x_2-x_1-1} (G_1)_{r_2,r_1}\left(\fw\right) \notag\\
 &=
 \frac{1}{(2\pi i)^2}
 \int_{C_{R_2}} \frac{dz_1}{z_1^{x_1+1}} \int_{C_{R_1}} dz_2 z_2^{x_2-1} 
 \tilde{\Ig}_1(r_1,z_1;r_2,z_2) 
 +
 G_1(r_1,x_1;r_2,x_2).
\end{align}
In the first equality the second term on the right hand side
appears from the contribution of a pole at $z_1=z_2$.
Hence one gets the double contour integral representation for $I_1$,
\begin{equation}
 I_1(r_1,x_1;r_2,x_2)
 =
 \frac{1}{(2\pi i)^2}
 \int_{C_{R_2}} \frac{dz_1}{z_1^{x_1+1}} \int_{C_{R_1}} dz_2 z_2^{x_2-1} 
 \tilde{\Ig}_1(r_1,z_1;r_2,z_2) .
\end{equation}

Let us first consider the asymptotics of $\tilde{S}_1$.
Since we can follow the discussions in \cite{Jo2002p} for a large part,
we omit details and state the main steps. If we write
\begin{equation}
\label{defg}
g_{\mu,\b}(z)
=
(1+\b)\log(z-\a)-(1-\b)\log(1-\a z)-(\mu+\b)\log z,
\end{equation}
we have
\begin{align}
 \label{Sa}
 &\quad
 \tilde{S}_1(r_1=2u_1,x_1;r_2=2u_2,x_2) \notag\\
 &=
 \frac{(1-\a)^{2(u_2-u_1)}}{(2\pi i)^2}
 \int_{C_{R_1}} \frac{dz_1}{z_1} \int_{C_{R_2}} \frac{dz_2}{z_2}
 \frac{z_2^{x_2-N(\mu_2-1)}}{z_1^{x_1-N(\mu_1-1)}}
 e^{N(g_{\mu_1,\b_1}(z_1)+g_{\mu_2,\b_2}(1/z_2))} \notag\\
 &\quad\times
 \frac{z_1}{z_1-z_2}
 \left\{1 + \frac{z_1-z_2}{(1+z_1)(z_2-1) } \right\},
\end{align}
where $\b_1=u_1/N,\b_2=-u_2/N$.  $\mu_1,\mu_2$ are arbitrary
constants at this stage. 
We would like to apply the saddle point method to this integral.
In general for each value of $\mu$, there are two critical points
$z_c^{\pm}=p(\mu,\b)\pm\sqrt{p(\mu,\b)^2-q(\mu,\b)}$ with
\begin{align}
 p(\mu,\b) &= \frac{\mu(1+\a^2)-(1-\a^2)}{2\a(\mu-\b)}, \\
 q(\mu,\b) &= \frac{\mu+\b}{\mu-\b}. 
\end{align}
Since we are considering the scaling in (\ref{scaledH}), we will set 
\begin{equation}
 x_i = N a(\b_i)  + d(\b_0) N^{\frac13} \xi_i,
\end{equation}
with $\xi_i$ the scaled space variable. 
Accordingly we set $\mu_i = \mu_c(\b_i)$ with
\begin{equation}
\mu_c(\b) = a(\b)+1=
\frac{1}{1-\a^2}\left(1+\a^2+2\a\sqrt{1-\b^2}\right).
\end{equation}
For this special value of $\mu$, the two critical points, 
$z_c^{\pm}$, merge to the double critical point $p_c(\b)$,
where $g'_{\mu_c(\b),\b}(p_c(\b)) = g''_{\mu_c(\b),\b}(p_c(\b))=0$, 
given by
\begin{equation}
 \label{pcdef}
 p_c(\b)
 =
 p(\mu_c(\b),\b)
% =
% \frac{2\a+(1+\a^2)\sqrt{1-\b^2}}{1+\a^2+2\a\sqrt{1-\b^2}-\b(1-\a^2)}\notag\\
 =
 \frac{\sqrt{1+\b}+\a \sqrt{1-\b}}{\sqrt{1-\b}+\a\sqrt{1+\b}}. 
\end{equation}

The main contribution to the integral in (\ref{Sa}) comes from around 
the double critical points $z_1 \sim p_c(\b_1),z_2\sim 1/p_c(\b_2)$.
The paths of integration may be deformed to 
\begin{align}
 \label{pathz1}
 z_1           &= p_c(\b_1)\left(1-\frac{i}{d(\b_0)N^{1/3}}w_1\right), \\
 \label{pathz2}
 \frac{1}{z_2} &= p_c(\b_2)\left(1-\frac{i}{d(\b_0)N^{1/3}}w_2\right), 
\end{align}
where $w_i = \z_i+i\eta_i$ with $-\i<\z_i<\i$ ($i=1,2$) 
and $\eta_1,\eta_2 >0$ are fixed.
For a fixed $\b_0$, we take
\begin{equation}
 r_i = 2 u_i = 2N \left(\b_0+\frac{c(\b_0)\t_i}{N^{1/3}}\right),
\end{equation}
where $\tau_i$ is the scaled time variable.
To leading order we have
\begin{align}
\label{pc1}
 p_c(\b_1) &= p_c(\b_0)+p_c'(\b_0)(\b_1-\b_0) 
           = p_c(\b_0)\left(1+\frac{\t_1}{d(\b_0)N^{1/3}}\right), \\
\label{pc2}
 p_c(\b_2) &= p_c(-\b_0)+p_c'(-\b_0)(\b_2+\b_0) 
           = p_c(-\b_0)\left(1-\frac{\t_2}{d(\b_0)N^{1/3}}\right),
\end{align}
where
\begin{equation}
 c(\b_0)
 =
 \frac{p_c(\b_0)}{p_c'(\b_0) d(\b_0)}
 =
 \frac{p_c(-\b_0)}{p_c'(-\b_0) d(\b_0)}
\end{equation}
is used. 

Expanding around $z_1 \sim p_c(\b_1)$ to order $O(1)$, one finds
\begin{align}
 N g_{\mu_c(\b_1),\b_1}(z_1)
 &\sim
 N g_{\mu_c(\b_1),\b_1}(p_c(\b_1))
 +
 \frac16 g_{\mu_c(\b_1),\b_1}'''(p_c(\b_1)) 
 \left(-i\frac{p_c(\b_1)}{d(\b_0)}w_1\right)^3 \notag\\
 &\sim
 N g_{\mu_c(\b_1),\b_1}(p_c(\b_1))
 +
 \frac{i}{3}\frac{p_c(\b_0)^3}{2 d(\b_0)^3} 
 g_{\mu_c(\b_0),\b_0}'''(p_c(\b_0)) w_1^3 \notag\\
 &=
 N g_{\mu_c(\b_1),\b_1}(p_c(\b_1)) 
 +
 \frac{i}{3}w_1^3
\label{Ngex}
\end{align}
after some computation.
%Using
%\begin{align}
% p_c(\b)-\a
% &=
% \frac{(1-\a^2)\sqrt{1+\b}}{\sqrt{1-\b}+\a\sqrt{1+\b}}, \\
% 1-\a p_c(\b)
% &=
% \frac{(1-\a^2)\sqrt{1-\b}}{\sqrt{1-\b}+\a\sqrt{1+\b}}, \\ 
% \mu_c(\b)+\b
% &=
% \frac{(\sqrt{1+\b}+\a\sqrt{1-\b})^2}{1-\a^2},
%\end{align}
%we find
%\begin{equation}
% g_{\mu_c(\b_0),\b_0}'''(p_c(\b_0)) 
% =
% \frac{2}{p_c(\b)^3} d(\b_0)^3. 
%\end{equation}
%Hence one has
%\begin{equation}
% N g_{\mu_c(\b_1),\b_1}(z_1)
% \sim
% N g_{\mu_c(\b_1),\b_1}(p_c(\b_1)) 
% +
% \frac{i}{3}w_1^3. 
%\end{equation}
Similarly one gets
\begin{equation}
 N g_{\mu_c(\b_2),\b_2}(1/z_2)
 \sim
 N g_{\mu_c(\b_2),\b_2}(p_c(\b_2)) 
 +
 \frac{i}{3}w_2^3 . 
\end{equation}
Let us denote
\begin{align}
 \l_0(\b) 
 &= 
 g_{\mu_c(\b),\b}(p_c(\b)) \notag\\
 &=
 2\b\log(1-\a^2) + \frac12 (1+\b)\log(1+\b)-\frac12 (1-\b)\log(1-\b)
 \notag\\
 &\quad+
 \frac{1}{1-\a^2}(\sqrt{1-\b}+\a\sqrt{1+\b})^2
 \log(\sqrt{1-\b}+\a\sqrt{1+\b}) 
 \notag\\
 &\quad
 -
 \frac{1}{1-\a^2}(\sqrt{1+\b}+\a\sqrt{1-\b})^2
 \log(\sqrt{1+\b}+\a\sqrt{1-\b}).
\end{align}
Clearly, this is an odd function.
Therefore if we expand as
\begin{equation}
\label{Gexpa}
 \l(\b) 
 \sim
 \l_0(\b_0)+\l_1(\b_0)(\b-\b_0)+\l_2(\b_0)(\b-\b_0)^2+\l_3(\b_0)(\b-\b_0)^3
 + \cdots
\end{equation}
we have $\l_i(\b_0)=(-1)^{i+1}\l_i(-\b_0)$ for $i=0,1,2,\cdots$.
Expanding to order $O(1)$, we have
\begin{align}
 \label{Gb1}
 N g_{\mu_c(\b_1),\b_1}(p_c(\b_1))
 &\sim
 N\l_0(\b_0)+ \l_1(\b_0)c(\b_0)N^{2/3}\t_1 \notag\\
 &\quad +\l_2(\b_0)c(\b_0)^2 N^{1/3}\t_1^2 +\l_3(\b_0)c(\b_0)^3 \t_1^3, \\
 \label{Gb2}
 N g_{\mu_c(\b_2),\b_2}(p_c(\b_2))
 &\sim
 -N \l_0(\b_0)-\l_1(\b_0)c(\b_0)N^{2/3}\t_2 \notag\\
 &\quad
 -\l_2(\b_0)c(\b_0)^2 N^{1/3}\t_2^2 -\l_3(\b_0)c(\b_0)^3 \t_2^3. 
\end{align}

Combining (\ref{pathz1}),(\ref{pathz2}) with (\ref{pc1}),(\ref{pc2}) gives
\begin{align}
\label{z1pcS}
 z_1           
 &= 
 p_c(\b_0)\left(1 + \frac{1}{d(\b_0)N^{1/3}}(\t_1-iw_1)\right), \\
\label{z2pcS}
 \frac{1}{z_2} 
 &= 
 p_c(-\b_0)\left(1 - \frac{1}{d(\b_0)N^{1/3}}(\t_2+iw_2)\right),
\end{align}
so that 
\begin{equation}
 \frac{z_2^{x_2+N(1-\mu_c(\b_2))}}{z_1^{x_1+N(1-\mu_c(\b_1))}}
 \sim
 (p_c(\b_0))^{(\xi_2-\xi_1)d(\b_0) N^{1/3}}
 e^{\xi_2 \t_2-\xi_1 \t_1+i\xi_1 w_1+i\xi_2 w_2} .
\end{equation}
In addition, to leading order we have
\begin{equation}
 \label{z1z2asy}
 \frac{z_1}{z_1-z_2}
 \left\{1 + \frac{z_1-z_2}{(1+z_1)(z_2-1)} \right\}
 \sim
 -\frac{d(\b_0)N^{1/3}}{\t_2-\t_1+i(w_1+w_2)}, 
\end{equation}
to which the second term on the left hand side does not contribute.

Finally one obtains
\begin{align}
\label{Sasy}
 \tilde{S}_1
 &\sim
 (1-\a)^{2(u_2-u_1)} (p_c(\b_0))^{(\xi_2-\xi_1)d(\b_0)N^{1/3}}
 \frac{1}{d(\b_0)N^{1/3}} \notag\\
 &\quad
 e^{ \l_1(\b_0)c(\b_0)N^{2/3}(\t_1-\t_2)
    +\l_2(\b_0)c(\b_0)^2 N^{1/3}(\t_1^2-\t_2^2)
    +\l_3(\b_0)c(\b_0)^3 (\t_1^3-\t_2^3)+\xi_2 \t_2-\xi_1 \t_1} \notag\\
 &\quad
 \frac{1}{4\pi^2}\int_{\text{Im}w_1=\eta_1} dw_1 
                 \int_{\text{Im}w_2=\eta_2} dw_2 
 \left(-\frac{1}{\tau_2-\tau_1+i(w_1+w_2)}\right)
 e^{i\xi_1 w_1 + i\xi_2 w_2 + \frac{i}{3}(w_1^3+w_2^3)}.
\end{align}
We also need the asymptotics of 
\begin{align}
 &\quad
 \p_{r_1,r_2}(x_1,x_2) \notag\\
 &=
 \frac{(1-\a)^{2(u_2-u_1)}}{2\pi i}
 \int_{C_1} \frac{dz}{z}
 z^{x_2-x_1} [(1-\a z)(1-\a/z)]^{u_2-u_1} \notag\\
 &=
 \frac{(1-\a)^{2(u_2-u_1)}}{2\pi i}
 \int_{C_1} \frac{dz}{z}
 z^{x_2-N(\mu_c(\b_2)-1)-x_1+N(\mu_c(\b_1)-1) }
 e^{N g_{\mu_c(\b_1),\b_1}(z) + N g_{\mu_c(\b_2),\b_2}(1/z)}.
\end{align}
Let us set
\begin{equation}
 z = p_c(\b_0)\left(1+\frac{i\sigma}{d(\b_0) N^{1/3}}\right) . 
\end{equation}
From (\ref{pc1}), this can be rewritten as 
\begin{equation}
z \sim p_c(\b_1)\left(1-\frac{1}{d(\b_0)N^{1/3}}(\t_1-i\sigma)  \right).
\end{equation}
Expansion around $z\sim p_c(\b_1)$ as in (\ref{Ngex}) gives
\begin{equation}
 N g_{\mu_c(\b_1),\b_1}(z)
 \sim
 N g_{\mu_c(\b_1),\b_1}(p_c(\b_1)) -\frac13 (\t_1-i\sigma)^3 .
\end{equation}
Similarly
\begin{equation}
\frac{1}{z} \sim p_c(\b_2)\left(1+\frac{1}{d(\b_0)N^{1/3}}(\t_2-i\sigma)\right)
\end{equation}
leads to 
\begin{equation}
 N g_{\mu_c(\b_2),\b_2}\left(\frac{1}{z}\right)
 \sim
 N g_{\mu_c(\b_2),\b_2}(p_c(\b_2)) + \frac13 (\t_2-i\sigma)^3 .
\end{equation}
One also finds
\begin{equation}
 z^{x_2-N(\mu_c(\b_2)-1)-x_1+N(\mu_c(\b_1)-1)} 
 \sim
 (p_c(\b_0))^{d(\b_0)N^{1/3}(\xi_2-\xi_1)}
 e^{i\sigma (\xi_2-\xi_1)}.
\end{equation}
Hence one gets
\begin{align}
 \p_{r_1,r_2}(x_1,x_2)
 &\sim
 (1-\a)^{2(u_2-u_1)} (p_c(\b_0))^{(\xi_2-\xi_1)d(\b_0)N^{1/3}}
 \frac{1}{d(\b_0)N^{1/3}} \notag\\
 &\quad
 e^{ \l_1(\b_0)c(\b_0)N^{2/3}(\t_1-\t_2)
    +\l_2(\b_0)c(\b_0)^2 N^{1/3}(\t_1^2-\t_2^2)
    +\l_3(\b_0)c(\b_0)^3 (\t_1^3-\t_2^3)-\frac{\t_1^3}{3}+\frac{\t_2^3}{3}} 
 \notag\\
 &\quad
 \frac{1}{2\pi}\int_{-\i}^{\i} d\sigma
 e^{i(\xi_2-\xi_1+\t_1^2-\t_2^2)\sigma-(\t_2-\t_1)\sigma^2}.
\label{pasy}
\end{align}

For $I_1$, one has
\begin{align}
 &\quad 
 I_1(r_1=2u_1,x_1;r_2=2u_2,x_2) \notag\\
 &=
 \frac{(1-\a)^{2(u_1+u_2)}}{(2\pi i)^2}
 \int_{C_{R_2}} \frac{dz_1}{z_1} \int_{C_{R_1}} \frac{dz_2}{z_2}
 \frac{z_2^{x_2-N(\mu_c(\b_2)-1)}}{z_1^{x_1-N(\mu_c(\b_1)-1)}}
 e^{N(g_{\mu_c(\b_1),\b_1}(z_1)+g_{\mu_c(\b_2),\b_2}(1/z_2))} \notag\\
 &\quad
 \times\frac{z_1}{z_1-z_2}
 \left\{ \frac12\frac{z_2+1}{z_2-1}-\frac12\frac{1+z_1}{1-z_1} \right\},
\end{align}
where $\b_i=-u_i/N$. The critical points of $z_1,z_2$ are the 
again given by $z_1=p_c(\b_1),z_2=1/p_c(\b_2)$.
To leading order, the paths are 
\begin{align}
\label{z1pcI}
 z_1           
 &= 
 p_c(-\b_0)\left(1 - \frac{1}{d(\b_0)N^{1/3}}(\t_1+iw_1)\right), \\
\label{z2pcI}
 \frac{1}{z_2} 
 &= 
 p_c(-\b_0)\left(1 - \frac{1}{d(\b_0)N^{1/3}}(\t_2+iw_2)\right),
\end{align}
so that we have
\begin{align}
\label{Iasy}
 I_1
 &\sim
 (1-\a)^{2(u_1+u_2)} \frac{p_c(-\b_0)^3}{(1+p_c(-\b_0))(1-p_c(-\b_0))^3}
 (p_c(\b_0))^{(\xi_1+\xi_2)d(\b_0)N^{1/3}}
 \frac{1}{d(\b_0)^3 N} \notag\\
 &\quad
 e^{ -2N\l_0(\b_0)
    -\l_1(\b_0)c(\b_0)N^{2/3}(\t_1+\t_2)
    -\l_2(\b_0)c(\b_0)^2 N^{1/3}(\t_1^2+\t_2^2)
    -\l_3(\b_0)c(\b_0)^3 (\t_1^3+\t_2^3)+\xi_1\t_1+\xi_2\t_2} \notag\\
 &\quad
 \frac{1}{4\pi^2}\int_{\text{Im}w_1=\eta_1} dw_1 
                 \int_{\text{Im}w_2=\eta_2} dw_2 
 \left(\tau_1-\tau_2+i(w_1-w_2)\right)
 e^{i\xi_1 w_1 + i\xi_2 w_2 + \frac{i}{3}(w_1^3+w_2^3)}. 
\end{align}

Similarly for $D_1$ one has
\begin{align}
 &\quad
 D_1(r_1=2u_1,x_1;r_2=2u_2,x_2) \notag\\
 &=
 \frac{1}{(2\pi i)^2 (1-\a)^{2(u_1+u_2)}}
 \int_{C_{R_1}} \frac{dz_1}{z_1} \int_{C_{R_2}} \frac{dz_2}{z_2}
 \frac{z_2^{x_2-N(\mu_c(\b_2)-1)}}{z_1^{x_1-N(\mu_c(\b_1)-1)}}
 e^{N(g_{\mu_c(\b_1),\b_1}(z_1)+g_{\mu_c(\b_2),\b_2}(1/z_2))} \notag\\
 &\quad
 \times\frac{z_1}{z_1-z_2}
 \left\{ \frac12\frac{z_1-1}{z_1+1}-\frac12\frac{1-z_2}{1+z_2} \right\},
\end{align}
where $\b_i=u_i/N$.
The critical points of $z_1,z_2$ are 
$z_1=p_c(\b_1),z_2=1/p_c(\b_2)$.
To leading order, the paths are 
\begin{align}
\label{z1pcD}
 z_1           
 &= 
 p_c(\b_0)\left(1 + \frac{1}{d(\b_0)N^{1/3}}(\t_1-iw_1)\right), \\
\label{z2pcD}
 \frac{1}{z_2} 
 &= 
 p_c(\b_0)\left(1 + \frac{1}{d(\b_0)N^{1/3}}(\t_2-iw_2)\right),
\end{align}
so that we have
\begin{align}
\label{Dasy}
 D_1
 &\sim
 (1-\a)^{-2(u_1+u_2)} \frac{p_c(\b_0)}{(1+p_c(\b_0))(1-p_c(\b_0))^3}
 (p_c(\b_0))^{-(\xi_1+\xi_2)d(\b_0)N^{1/3}}
 \frac{1}{d(\b_0)^3 N} \notag\\
 &\quad
 e^{ 2N\l_0(\b_0)
    +\l_1(\b_0)c(\b_0)N^{2/3}(\t_1+\t_2)
    +\l_2(\b_0)c(\b_0)^2 N^{1/3}(\t_1^2+\t_2^2)
    +\l_3(\b_0)c(\b_0)^3 (\t_1^3+\t_2^3)-\xi_1\t_1-\xi_2\t_2} \notag\\
 &\quad
 \frac{1}{4\pi^2}\int_{\text{Im}w_1=\eta_1} dw_1 
                 \int_{\text{Im}w_2=\eta_2} dw_2 
 \left(\tau_1-\tau_2-i(w_1-w_2)\right)
 e^{i\xi_1 w_1 + i\xi_2 w_2 + \frac{i}{3}(w_1^3+w_2^3)}.
\end{align}

Now suppose that we substitute the asymptotic expressions 
(\ref{Sasy}),(\ref{pasy}),(\ref{Iasy}),(\ref{Dasy}) into (\ref{K1def}).
Notice that the kernel of the form
$\begin{bmatrix}  a & b \\ c & d \end{bmatrix}$ and
$\begin{bmatrix}  \a a & \b b \\ c/\b & d/\a \end{bmatrix}$ give the 
same value for the determinant and hence that
some of the prefactors in (\ref{Sasy}),(\ref{pasy}),
(\ref{Iasy}),(\ref{Dasy}) have no effect on the value of 
the determinant. Moreover, in our scaling limit, 
the off-diagonal elements, $I_1$ and $D_1$, vanish due to 
the difference of order in $N$.
The diagonal elements are transpose to each other and hence 
one finds
\begin{align}
 &\quad 
 \P[h(r_1,2N)\leq X_1,\cdots,h(r_m,2N)\leq X_m]
 = \sqrt{\det(1+K_1 g)} \notag\\
 &\to
 \sqrt{(\det(1+\K_2 \G))^2} 
 =
 \det(1+\K_2 \G),
\end{align}
where $g$ is $ g(r_i,x) = -\chi_{J_i}(x)$ ($i=1,2,\cdots, m$),
where
$J_i=(X_i,\i)$ and 
$r_i=2\b_0 N +2 c(\b_0) N^{2/3} \t_i,
X_i=a(\b_0 + \frac{c(\b_0)\t_i}{N^{1/3}}) N + s_i d(\b_0) N^{1/3}$
for $i=1,2,\cdots ,m$.
\qed

\subsection{Near the Origin}
\label{origin}
In the last subsection, we saw that the fluctuation in the 
bulk is described by the Airy process. On the other hand, 
as already mentioned, we know that the height 
fluctuation at the origin is given by the GOE Tracy-Widom 
distribution. In this subsection, we are interested in the 
crossover between them.
Let us define the scaled height variable near the origin as
\begin{equation}
\label{scaledHo}
 H_N(\t)=\frac{h(r=2cN^{2/3}\t,t=M=2N)-aN}{d N^{1/3}} +\t^2 ,
\end{equation}
where
\begin{align}
 a &= a(0) =\frac{2\a}{1-\a},\\
 d &= d(0) =\frac{\a^{1/3}(1+\a)^{1/3}}{1-\a}, \\
 c &= c(0) =\frac{(1+\a)^{\frac23}}{\a^{\frac13}}.
\end{align}
The second term in (\ref{scaledHo}) comes from the expansion 
of $a(\b)$ in (\ref{defa}) around $\b=0$.

We show that the fluctuation of the model near the origin
is described by the process which gives the orthogonal-unitary 
transition in random matrix theory.
Namely, we show
\begin{theorem}
\begin{equation}
\label{detKlim}
 \lim_{N\to\i} \P[H_N(\t_1) \leq s_1, \cdots , H_N(\t_m) \leq s_m]
 =
 \sqrt{\det(1+\K_1 \G)},
\end{equation}
where $\G(\t_j,\xi)=-\chi_{(s_j,\i)}(\xi)$ ($j=1,2,\cdots,m$) and 
$\K_1$ is the $2\times 2$ matrix kernel, for which a representative 
element is given by 
\begin{equation}
 \K_1(\t_1,\xi_1;\t_2,\xi_2)
 =
 \begin{bmatrix}
  \S_1(\t_1,\xi_1;\t_2,\xi_2) & \D_1(\t_1,\xi_1;\t_2,\xi_2) \\
  \I_1(\t_1,\xi_1;\t_2,\xi_2) & \S_1(\t_2,\xi_2;\t_1,\xi_1) 
 \end{bmatrix},
\end{equation}
with the matrix elements being
\begin{align}
 \label{defSlim}
 \S_1(\t_1,\xi_1;\t_2,\xi_2)
 &=
 \tilde{\S_1}(\t_1,\xi_1;\t_2,\xi_2)-\Phi_{\t_1,\t_2}(\xi_1,\xi_2), \\
 \label{defStlim}
 \tilde{\S_1}(\t_1,\xi_1;\t_2,\xi_2)
 &=
 \int_0^{\i} d\l e^{-\l(\t_1-\t_2)} \Ai(\xi_1+\l) \Ai(\xi_2+\l)
 +
 \frac12 \Ai(\xi_1) \int_0^{\i} d\l e^{-\l \t_2} \Ai(\xi_2-\l),\\
 \label{defIlim}
 \I_1(\t_1,\xi_1;\t_2,\xi_2)
 &=
 -\int_0^{\i} d\l e^{-\l\t_1} \Ai(\xi_1-\l) 
  \int_{\l}^{\i} dv e^{-v\t_2} \Ai(\xi_2-v)   \notag\\
 &\quad + 
  \int_0^{\i} d\l e^{-\l\t_2} \Ai(\xi_2-\l) 
   \int_{\l}^{\i} dv e^{-v\t_1} \Ai(\xi_1-v), 
 \\
 \label{defDlim}
 \D_1(\t_1,\xi_1;\t_2,\xi_2)
 &=
 -\frac14 \int_0^{\i} d\l e^{-\l \t_2} \Ai(\xi_2+\l) 
  \frac{d}{d \l} \left\{e^{-\l\t_1} \Ai(\xi_1+\l)\right\} \notag\\
 &\quad + 
 \frac14 \int_0^{\i} d\l e^{-\l \t_1} \Ai(\xi_1+\l) 
 \frac{d}{d \l} \left\{e^{-\l\t_2} \Ai(\xi_2+\l)\right\}.
\end{align}
Definition of $\Phi_{\t_1,\t_2}(\xi_1,\xi_2)$ is already given in 
(\ref{Phidef}).
\end{theorem}

\vspace{3mm}\noindent
{\bf Remark.} 
The same kernel appeared in the context of the orthogonal-unitary
transition in \cite{FNH1999} and the problem of vicious walks 
in \cite{NKT2003}.
For the special case of a fluctuation at $\t=0$, we have
\begin{equation}
 \lim_{N\to\i} \P[H_N(0) \leq s] = F_1(s),
\end{equation} 
where $F_1(s)$ is the GOE Tracy-Widom distribution \cite{TW1996}. 
It is remarked that this result was anticipated in \cite{Jo2000}
based on the Meixner orthogonal ensemble representation and was shown 
in \cite{BR2001c}.
On the other hand, for $\t\to\i$, $\I_1$ and $\D_1$ goes to zero.
We recover the Airy process, which is consistent with 
the results of the previous subsection.

\vspace{3mm}\noindent
{\bf Proof.} 
Let us derive (\ref{detKlim}).
We first notice that (\ref{defStlim}), (\ref{defIlim}),
(\ref{defDlim}) have double contour integral representation.
Assume $\t_1,\t_2 \geq 0$ and $\eta_1,\eta_2>0$.
For $\tilde{\S}_1$, one has
\begin{align}
 &\quad \tilde{\S}_1(\t_1,\xi_1;\t_2,\xi_2) \notag\\
 &=
 \frac{1}{4\pi^2}\int_{\text{Im}w_1=\eta_1} dw_1 
                 \int_{\text{Im}w_2=\eta_2} dw_2 
 \left(-\frac{1}{\tau_2-\tau_1+i(w_1+w_2)}+\frac{1}{2(\tau_2+iw_2)}\right)
 \notag\\
 &\quad
 e^{i\xi_1 w_1 + i\xi_2 w_2 + \frac{i}{3}(w_1^3+w_2^3)},
\end{align}
where $\eta_1+\eta_2+\t_1-\t_2>0$ and $\t_2-\eta_2>0$.
For $\I_1$, 
\begin{align}
 &\quad \I_1(\t_1,\xi_1;\t_2,\xi_2) \notag\\
 &=
 \frac{1}{4\pi^2}\int_{\text{Im}w_1=\eta_1} dw_1 
                 \int_{\text{Im}w_2=\eta_2} dw_2 
 \frac{-1}{(\tau_2+iw_2)(\tau_1+\tau_2+i(w_1+w_2))}
 e^{i\xi_1 w_1 + i\xi_2 w_2 + \frac{i}{3}(w_1^3+w_2^3)} \notag\\
 &-  
 (\t_1,\xi_1 \leftrightarrow \t_2,\xi_2),
\end{align}
where $\t_1-\eta_1>0,\t_2-\eta_2>0$.
For $\D_1$, 
\begin{align}
 &\quad \D_1(\t_1,\xi_1;\t_2,\xi_2) \notag\\
 &=
 \frac{1}{4\pi^2}\int_{\text{Im}w_1=\eta_1} dw_1 
                 \int_{\text{Im}w_2=\eta_2} dw_2 
 \frac{\tau_1-iw_1}{4(\tau_1+\tau_2-i(w_1+w_2))}
 e^{i\xi_1 w_1 + i\xi_2 w_2 + \frac{i}{3}(w_1^3+w_2^3)} \notag\\
 &-  
 (\t_1,\xi_1 \leftrightarrow \t_2,\xi_2),
\end{align}
where there is no additional condition on $\t_1,\t_2,\eta_1,\eta_1$.
These are easily obtained by using the integral representation of 
the Airy function,
\begin{equation}
\label{Aiint}
 \Ai(\xi) 
 =
 \frac{1}{2\pi} \int_{\text{Im}~ w=\eta} dw e^{i\xi w+i\frac{w^3}{3}}.
\end{equation}
In the following, we show, as $N\to\i$,
\begin{align}
 \label{Slim}
 S_1(r_1,x_1;r_2,x_2)
 &\sim
 e^{(\t_1^3-\t_2^3)/3-\xi_1\t_1+\xi_2\t_2}dN^{1/3} 
 \S_1(\t_1,\xi_1;\t_2,\xi_2) , \\
 \label{Ilim}
 I_1(r_1,x_1;r_2,x_2)
 &\sim
 e^{-(\t_1^3+\t_2^3)/3+\xi_1\t_1+\xi_2\t_2}
 \I_1(\t_1,\xi_1;\t_2,\xi_2) , \\
 \label{Dlim}
 D_1(r_1,x_1;r_2,x_2)
 &\sim
 e^{(\t_1^3+\t_2^3)/3-\xi_1\t_1-\xi_2\t_2}d^2N^{2/3}  
 \D_1(\t_1,\xi_1;\t_2,\xi_2),
\end{align}
where we set $r_i=2cN^{2/3}\t_i,x_i = a N +(\xi_i-\t_i^2)dN^{1/3}$
for $i=1,2$.
From these, (\ref{detKlim}) is almost obvious.

Let us first consider the asymptotics of $\tilde{S}_1$.
Basically one can follow the same route as in the last subsection
with $\b_0=0$, but there appear some simplifications and differences. 
For instance, (\ref{z1pcS}),(\ref{z2pcS}) reduce to 
\begin{equation}
 z_1           = 1 + \frac{1}{dN^{1/3}}(\t_1-iw_1), \quad
 \frac{1}{z_2} = 1 - \frac{1}{dN^{1/3}}(\t_2+iw_2).
\end{equation}
Then we have
\begin{equation}
 \frac{z_1}{z_1-z_2}
 \left\{ 1 + \frac{z_1-z_2}{(1+z_1)(z_2-1)} \right\}
 \sim
 -\frac{dN^{1/3}}{\t_2-\t_1+i(w_1+w_2)} 
 +\frac12\frac{dN^{1/3}}{\t_2+iw_2}.
\end{equation}
Compare this with (\ref{z1z2asy}), where 
the second term on the left hand side was negligible.

Expansion of $G(\b)$ in (\ref{Gexpa}) is now simply
\begin{equation}
 G(\b) = 2\log(1-\a)~ \b + \frac{\a}{3(1+\a)^2} \b^3,
\end{equation}
so that one has
\begin{align}
 e^{N g_{\mu_c(\b_1),\b_1}(p_c(\b_1))}
 &\sim
 (1-\a)^{2u_1} e^{\t_1^3/3}, \\
 e^{N g_{\mu_c(\b_2),\b_2}(p_c(\b_2))}
 &\sim
 (1-\a)^{-2u_2} e^{-\t_2^3/3},
\end{align}
instead of (\ref{Gb1}),(\ref{Gb2}).
Asymptotics of $\p$, which is already given in \cite{Jo2002p},
can be obtained by setting $\b_0=0$ in (\ref{pasy});
\begin{equation}
 \phi_{r_1,r_2}(x_1,x_2)
 =
 e^{(\t_2^3-\t_1^3)/3+\xi_1\t_1-\xi_2\t_2}dN^{1/3} 
 \Phi_{\t_1,\t_2}(\xi_1,\xi_2). 
\end{equation}
Combining these, we see (\ref{Slim}) holds.

As for $I_1$, (\ref{z1pcI}),(\ref{z2pcI}) reduce to
\begin{equation}
 z_1           = 1 - \frac{1}{dN^{1/3}}(\t_1+iw_1), \quad
 \frac{1}{z_2} = 1 - \frac{1}{dN^{1/3}}(\t_2+iw_2). 
\end{equation}
Since
\begin{equation}
 \frac{z_1}{z_1-z_2} 
 \left\{ \frac12\frac{z_2+1}{z_2-1}-\frac12\frac{1+z_1}{1-z_1} \right\}
 \sim
 \frac{d^2N^{2/3}}{\t_1+\t_2+i(w_1+w_2)}
 \left( \frac{-1}{\t_2+iw_2}
       -\frac{-1}{\t_1+iw_1}\right),  
\end{equation}
we see that (\ref{Ilim}) holds.

Finally, for $D_1$, 
(\ref{z1pcD}),(\ref{z2pcD}) reduce to
\begin{equation}
 z_1           = 1 + \frac{1}{dN^{1/3}}(\t_1-iw_1), \quad
 \frac{1}{z_2} = 1 + \frac{1}{dN^{1/3}}(\t_2-iw_2 ).
\end{equation}
Since
\begin{equation}
 \frac{z_1}{z_1-z_2} 
 \left\{ \frac12\frac{z_1-1}{z_1+1}-\frac12\frac{1-z_2}{1+z_2} \right\}
 \sim
 \frac14\frac{(\t_1-iw_1)-(\t_2-iw_2)}
             {\t_1+\t_2-i(w_1+w_2)},
\end{equation}
we see that (\ref{Dlim}) holds.
\qed

\setcounter{equation}{0}
\setcounter{theorem}{1}
\section{Model without Source at the Origin}
\label{symplectic}
In this section we study the case where $\g=0$ in (\ref{wii}).
As already mentioned in the introduction, this case corresponds 
to the PNG model without the source at the origin.
The analysis proceeds in a fairy parallel way to the orthogonal case.
The main difference is that there is a new restriction on the height 
lines $h_i$; $h_{2i}(0,t)=h_{2i+1}(0,t)+1$ ($i=0,1,\cdots,N/2-1$)
at time $t=M=2N$. 
Here and hereafter we assume $N$ is even.
This condition might look somewhat strange at first sight, but
in fact can be understood after a short reflection. 
An example of a multi-layer 
heights is given in Fig. 3, in which one sees that 
$h_{2i}$ and $h_{2i+1}$ form a pair at $r=0$.
As in the orthogonal case, we change the way of looking at 
multi-layer heights. 
The space coordinate $r$ will be interpreted as the time coordinate.
The height coordinate will be interpreted as the space coordinate
and is represented as $x$.
Then the restrictions read $x_i^M=1-i$ for $i=1,2,\cdots, N$
and $x_{2i}^0=x_{2i-1}^0-1$ for $i=1,2,\cdots,N/2$.

\subsection{Determinantal Process and Kernel}
We consider the weight of non-intersecting paths of 
$n$ particles given by 
\begin{equation}
 w_{n,M}(\bar{x}) 
 = 
 \det\left(\begin{array}{@{\,}c@{\,}}
\p_{0,1}(x_{2i-1}^0,x_j^{1})\\
\p_{0,1}(x_{2i-1}^0-1,x_j^{1})
\end{array}
\right)_{i=1,\cdots,\frac{n}{2}, j=1,\cdots ,n}\\ 
\prod_{r=1}^{M-1} \det (\p_{r,r+1}(x_i^r,x_j^{r+1}))_{i,j=1}^n,
\end{equation}
where $n$ is assumed to be even and $x_i^M$ ($i=1,2,\cdots,n$) is fixed. 
As in the orthogonal case, this is slightly different from the 
weight of the multi-layer PNG model even if we take $n=N$, 
but we employ this for the 
moment and take the infinite particles limit when it becomes easy.
The partition function and the probability of the non-intersecting 
paths are given by (\ref{par_fun}),(\ref{pNM}) respectively.
Again, our main focus is on the special case in (\ref{phie}),(\ref{phio}), 
but the results of this section do not depend on a specific choice of
$\p$.
For instance, when particles can hop only to nearest neighbor
sites for each time, the determinantal process is a variant 
of the vicious walks with a condition that walkers start in pairs
and end at neighboring sites \cite{Ka2003u}. 

Under this arrangement, we can show 
\begin{proposition}
\begin{equation}
 \sum_{\bar{x}} \prod_{r=0}^{M-1}\prod_{j=1}^n (1+g(r,x_j^r)) p_{n,M}(\bar{x})
 =
 \sqrt{\det(1+K_4 g)},
\label{detK4}
\end{equation}
where $g$ is some function.
The determinant on the right hand side is the Fredholm determinant,
where $K_4$ is a $2\times 2$ matrix kernel,
\begin{equation}
 K_4(r_1,x_1;r_2,x_2)
 =
 \begin{bmatrix}
  S_4(r_1,x_1;r_2,x_2) & D_4(r_1,x_1;r_2,x_2) \\
  I_4(r_1,x_1;r_2,x_2) & S_4(r_2,x_2;r_1,x_1) 
 \end{bmatrix},
\end{equation}
with the matrix elements being
\begin{align}
\label{S4fin}
 S_4(r_1,x_1;r_2,x_2)
 &=
 \tilde{S_4}(r_1,x_1;r_2,x_2) - \phi_{r_1,r_2}(x_1,x_2), \\
\label{St4fin}
 \tilde{S}_4(r_1,x_1;r_2,x_2) 
 &=
 -\sum_{i,j=1}^n \p_{r_1,M}(x_1,x_i^M) (A_4^{-1})_{i,j} G_4(r_2,x_2;M,x_j^M), \\
\label{I4fin}
 I_4(r_1,x_1;r_2,x_2)
 &=
 \tilde{I_4}(r_1,x_1;r_2,x_2)  -G_4(r_1,x_1;r_2,x_2) ,\\ 
\label{I4tfin}
 \tilde{I}_4(r_1,x_1;r_2,x_2)  
 &=
 -\sum_{i,j=1}^n G_4(r_1,x_1;M,x_i^M) 
 (A_4^{-1})_{i,j} G_4(r_2,x_2;M,x_j^M) ,\\
\label{D4fin}
 D_4(r_1,x_1;r_2,x_2)
 &=
 \sum_{i,j=1}^n  \p_{r_1,M}(x_1,x_i^M) (A_4^{-1})_{i,j} \p_{r_2,M}(x_2,x_j^M).
\end{align}
Here
\begin{equation}
\label{A4def}
 (A_4)_{ij} 
 = 
 \sum_{y_1,y_2} s(y_2-y_1) \p_{0,M}(y_1,x_i^M) \p_{0,M}(y_2,x_j^M),
\end{equation}
\begin{equation}
\label{G4def}
 G_4(r_1,x_1;r_2,x_2) 
 =
 \sum_{y_1,y_2} s(y_2-y_1) \p_{0,r_1}(y_1,x_1) \p_{0,r_2}(y_2,x_2),
\end{equation}
\begin{equation}
\label{sdet}
  s_{ij} =s(i-j) = \delta_{i+1,j}-\delta_{i,j+1}.
\end{equation}
\end{proposition}

\vspace{3mm}\noindent
{\bf Remark.} 
The subscript 4 refers to the fact that this case is related to 
the symplectic-unitary transition in random matrix theory.
It should also be remarked that at this stage the infinite particles 
limit is taken easily. One only replaces the summation,
$\sum_{i,j=1}^n$, in (\ref{St4fin}),(\ref{I4tfin}),(\ref{D4fin}) 
with $\sum_{i,j=1}^{\i}$.

\vspace{3mm}\noindent
{\bf Proof.} 
As in the orthogonal case, we derive (\ref{detK4}) by generalizing 
the methods of \cite{Jo2002p,TW1998}. 
Using the Heine identity (\ref{Heine}), the partition function 
$Z_{n,M}[g]$ is written as
\begin{equation}
 \label{SZmn}
  Z_{n,M}[g]=\frac{1}{(n!)^M} \sum_{x^0} 
\det\left(\begin{array}{@{\,}c@{\,}}
\p_{0,M}^g(x_{2i-1}^0,x_j^M)\\
\p_{0,M}^g(x_{2i-1}^0-1,x_j^M)
\end{array}\right)_{i=1,\cdots,n/2,j=1,\cdots, n}.
\end{equation}
Here
\begin{align}
 \label{Spg}
 \p^g_{0,M}(x^0_i, x^M_j)
 &=
 \sum_{X_1,\cdots,X_{M-1}} 
 (1+g(0,x_i^0))\phi_{0,1}(x_i^0,X_1)(1+g(1,X_1))\cdots  
 \phi_{M-1,M}(X_{M-1},x_j^M) \notag\\
 &=
 \phi_{0,M}(x_i^0,x_j^M)+(\phi_{0,r_1}\cdot g\psi_j)(x_i^0,x_j^M).
\end{align}
In the second equality we have used $\psi_j$ defined in (\ref{defpsi}).
From (\ref{SZmn}), (\ref{Spg}) and the identity \cite{TW1996,deB1955},
\begin{align}
 &\quad \left(\sum_{x_1,\cdots,x_n} \det(\phi_i(x_j)
 \psi_i(x_j))_{i=1\cdots 2n,j=1\cdots n}  \right)^2 \notag\\
 &=((2n)!)^2
 \det\left( \sum_{y}(\phi_i(y)\psi_j(y)-\phi_j(y)\psi_i(y))
     \right)_{i,j=1}^{2n} , 
 \label{eq:pretty}
\end{align}
one finds 
\begin{equation}
 Z_{n,M}[g]^2
 =
 \det \Bigl[
   (A_4)_{i,j} + (A_4^{(1)})_{i,j} + (A_4^{(2)})_{i,j} + (A_4^{(3)})_{i,j}
      \Bigr]_{i,j=1}^n,
\end{equation}
where $A_4$ is defined in (\ref{A4def}) and 
\begin{align}
 (A_4^{(1)})_{i,j}
 &= 
 \sum_{y}[(\phi_{0,r_1}\cdot g\psi_j)(y,x_j^M)\phi_{0,M}(y+1,x_k^M)
 -\phi_{0,M}(y,x_k^M)(\phi_{0,r_1}\cdot g\psi_j)(y+1,x_j^M)]\nonumber\\
 &=
 \sum_{r,x}g\psi_{i}\cdot (G_4)_j\\
 (A_{4}^{(2)})_{i,j}
 &=
 \sum_{y}[\phi_{0,M}(y,x_j^M)(\phi_{0,r_1}\cdot g\psi_k)(y+1,x_k^M)
 -(\phi_{0,r_1}\cdot g\psi_k)(y,x_k^M)\phi_{0,M}(y+1,x_j^M)]\nonumber\\
 &=
 -\sum_{r,x}g\psi_j\cdot (G_4)_i\\
 (A_{4}^{(3)})_{i,j}
 &=
 \sum_{y}[(\phi_{0,r_1}\cdot g\psi_j)(y,x_j^M)
 (\phi_{0,r_1}\cdot g\psi_k)(y+1,x_k^M) \notag\\
 &\quad
 -(\phi_{0,r_1}\cdot g\psi_k)(y,x_k^M)(\phi_{0,r_1}\cdot g\psi_j)(y+1,x_j^M)]
 \nonumber\\
 &=-\sum_{r,x}g\psi_j\cdot G_4(g\psi_i).
\end{align}
We should notice that the forms of $A_4^{(1)},A_4^{(2)}$ and $A_4^{(3)}$ are 
the same as the corresponding ones in the orthogonal case, (\ref{A11}),
(\ref{A12}) and (\ref{A13}), respectively. 
Only difference is that sgn in (\ref{Adef}) and (\ref{G1def}) 
is replaced by $s$, (\ref{sdet}), in (\ref{A4def}) and (\ref{G4def}). 
Thus we can calculate the kernel along the line in section 3.
The result is 
\begin{align}
 & \left(\frac{Z_{n,M}[g]}{Z_{n,M}[0]}\right)^2 \notag\\
 &=
 \det\left(1+
  \begin{bmatrix}
   -\sum_{i,j} \phi_i \otimes (A_4^{-1})_{i,j} (G_4)_j -\phi  
   & \sum_{i,j} \phi_i \otimes (A_4^{-1})_{i,j} \phi_j \\
   -\sum_{i,j} (G_4)_i \otimes (A_4^{-1})_{i,j} (G_4)_j  - G_4 
   & \sum_{i,j} (G_4)_i \otimes (A_4^{-1})_{i,j}\phi_j -\,^t \phi
  \end{bmatrix} g 
 \right) .
\end{align}
Recalling the definitions of $\phi_i$ and $(G_4)_i$, 
we see that (\ref{detK4}) holds.
\qed

\subsection{Scaling Limit}
\subsubsection{Generating Functions}
The symbol $a_4(z)$ of the matrix $A_4$ is computed as
\begin{equation}
  \label{eq:w}
  a_4(z)
 =f_{0,M}\left(\frac{1}{z}\right) s_4(z) f_{0,M}(z),
\end{equation}
where 
\begin{equation}
  s_4(z)=\sum_{k\in\Z}  s(k) z^k = -z+\frac{1}{z}.
\end{equation}
The winding number of $a_4(z)$ is again not zero, 
but the difficulty can be overcome as in the orthogonal case.
$s_4(z)$ can be expressed in two ways
\begin{equation}
 s_4(z) = s_{4+}^{(1)}(z) s_{4-}^{(1)}(z)  
        = s_{4+}^{(2)}(z) s_{4-}^{(2)}(z) ,
\end{equation}
with
\begin{align}
 s_{4+}^{(1)}(z) &= 1+z,  \quad s_{4-}^{(1)}(z) = -1+\frac{1}{z}, \\
 s_{4+}^{(2)}(z) &= 1-z,  \quad  s_{4-}^{(2)}(z) = 1+\frac{1}{z}.
\end{align}
Then in terms of $a_{4+}^{(i)},a_{4-}^{(i)}$ for $i=1,2$ defined by
\begin{align}
 a_{4+}^{(i)}(z) = f_{0,M;-}\left(\frac{1}{z}\right) 
                   f_{0,M;+}(z) s_{4+}^{(i)}(z), \\
 a_{4-}^{(i)}(z) = f_{0,M;+}\left(\frac{1}{z}\right) 
                   f_{0,M;-}(z) s_{4-}^{(i)}(z),
\end{align}
$A_4^{-1}$ is expressed as
\begin{equation}
 A_4^{-1}
 =
 \frac12\left\{ T\left(\frac{1}{a_{4+}^{(1)}}\right) 
                T\left(\frac{1}{a_{4-}^{(1)}}\right)
               +T\left(\frac{1}{a_{4+}^{(2)}}\right) 
                T\left(\frac{1}{a_{4-}^{(2)}}\right) 
        \right\}.
\end{equation}
Using the generating functions of $A_4^{-1}$ and $G_4$,
\begin{align}
&\quad
 \sum_{i,j=1}^{\i} z_1^{1-i} (A_4^{-1})_{i,j} z_2^{j-1} \notag\\
 &=
\label{zAz4}
 \frac{z_1}{z_1-z_2}
 \frac{1}{f_{0,M;-}(z_1)f_{0,M;+}(\fz)f_{0,M;+}(z_2)f_{0,M;-}(\fw)} 
 \frac12 \left\{ \frac{1}{s_{4+}^{(1)}(\fz)s_{4-}^{(1)}(\fw)}+
                 \frac{1}{s_{4+}^{(2)}(\fz)s_{4-}^{(2)}(\fw)} \right\},\notag\\
 &(G_4)_{r_1,r_2}(z)
 =
 -f_{0,r_1}(z) s_4(z) f_{0,r_2}\left(\frac{1}{z}\right),
\end{align}
we can calculate the generating functions of $\tilde{S}_4,\tilde{I}_4,
D_4$;
\begin{align}
 &\quad
 \tilde{\Sg}_4(r_1,z_1;r_2,z_2) 
 =
 \sum_{x_1,x_2\in\Z} \tilde{S}_4(r_1,x_1;r_2,x_2) z_1^{x_1} z_2^{-x_2} \notag\\
 &= 
 \frac{z_1}{z_1-z_2} \frac{f_{r_1,M;-}(\fz) f_{0,r_2;+}(\fw) f_{0,M;-}(z_2)}
                          {f_{0,M;-}(z_1) f_{0,r_1;+}(\fz) f_{r_2,M;-}(\fw)}
 \frac12\left\{ \frac{s_{4+}^{(1)}(\fw)}{s_{4+}^{(1)}(\fz)}
               +\frac{s_{4+}^{(2)}(\fw)}{s_{4+}^{(2)}(\fz)}\right\}, \notag\\
 &=
\label{Sgg4}
 \frac{(1-\a)^{2(u_2-u_1)} (1-\a/z_1)^{N+u_1} (1-\a z_2)^{N-u_2}}
      {(1-\a z_1)^{N-u_1} (1-\a/z_2)^{N+u_2}} 
 \frac{z_1}{z_1-z_2} 
 \left\{1 + \frac{z_1-z_2}{z_2(1-z_1^2)} \right\}, \\
&\quad\tilde{\Ig}_4(r_1,z_1;r_2,z_2) 
 = 
 \sum_{x_1,x_2\in\Z} \tilde{I}_4(r_1,x_1;r_2,x_2) z_1^{x_1} z_2^{-x_2} \notag\\
 &=
 \frac{z_1}{z_1-z_2} \frac{f_{0,r_1;+}(z_1) f_{0,M;-}(\fz) 
                           f_{0,r_2;+}(\fw) f_{0,M;-}(z_2)}
                          {f_{r_1,M;-}(z_1) f_{r_2,M;-}(\fw)}
 \notag\\
 &\quad \times
 \frac12\left\{ s_{4-}^{(1)}\left(\fz\right)s_{4+}^{(1)}\left(\fw\right)
               +s_{4-}^{(2)}\left(\fz\right)s_{4+}^{(2)}\left(\fw\right)\right\}, 
 \notag\\
 &=
\label{Igg4}
 \frac{(1-\a)^{2(u_1+u_2)} (1-\a/z_1)^{N-u_1} (1-\a z_2)^{N-u_2}}
      {(1-\a z_1)^{N+u_1} (1-\a/z_2)^{N+u_2}} 
 \frac{z_1}{z_1-z_2}
 \left\{-z_1+\frac{1}{z_2}\right\}, \\
 &\quad
 \Dg_4(r_1,z_1;r_2,z_2) 
 = 
 \sum_{x_1,x_2\in\Z} D_4(r_1,x_1;r_2,x_2) z_1^{x_1} z_2^{-x_2} \notag\\
 &=
 \frac{z_1}{z_1-z_2} \frac{f_{r_1,M;-}(\fz) f_{r_2,M;-}(z_2)}
                          {f_{0,M;-}(z_1)   f_{0,r_1;+}(\fz) 
                           f_{0,r_2;+}(z_2) f_{r_2,M;-}(\fw)}
 \frac12\left\{ \frac{1}{s_{4+}^{(1)}(\fz)s_{4-}^{(1)}(\fw)}
               +\frac{1}{s_{4+}^{(2)}(\fz)s_{4-}^{(2)}(\fw)}\right\} \notag\\
 &=
 \frac{(1-\a/z_1)^{N+u_1} (1-\a z_2)^{N+u_2}}
      {(1-\a)^{2(u_1+u_2)} (1-\a z_1)^{N-u_1} (1-\a/z_2)^{N-u_2}} 
 \notag\\ 
\label{Dgg4}
 &\quad \times
 \frac{-z_1}{z_1-z_2}
 \left\{ \frac12\frac{1}{z_1+1}\frac{1}{1-z_2}
 +\frac12\frac{1}{1-z_1}\frac{1}{1+z_2} \right\}.
\end{align}

\subsubsection{Bulk}
In the bulk region, the asymptotic behavior is just the same 
as in the orthogonal case.  
\begin{proposition}
For the scaled height variable defined 
by (\ref{scaledH}), (\ref{detKlim2}) holds in the symplectic case 
as well.  
\end{proposition}

\vspace{3mm}\noindent
{\bf Proof.} 
To show this, we need the asymptotics of $\tilde{S}_4,\tilde{I}_4$ and
$D_4$ in the bulk region. 
Applying the same reasoning as in 4.2, one finds
\begin{align}
\label{S4asy}
 \tilde{S}_4
 &\sim
 (1-\a)^{2(u_2-u_1)} (p_c(\b_0))^{(\xi_2-\xi_1)d(\b_0)N^{1/3}}
 \frac{1}{d(\b_0)N^{1/3}} \notag\\
 &\quad
 e^{ G_1(\b_0)c(\b_0)N^{2/3}(\t_1-\t_2)
    +G_2(\b_0)c(\b_0)^2 N^{1/3}(\t_1^2-\t_2^2)
    +G_3(\b_0)c(\b_0)^3 (\t_1^3-\t_2^3) +\xi_2\t_2-\xi_1\t_1} \notag\\
 &\quad
 \frac{1}{4\pi^2}\int_{\text{Im}w_1=\eta_1} dw_1 
                 \int_{\text{Im}w_2=\eta_2} dw_2 
 \left(-\frac{1}{\tau_2-\tau_1+i(w_1+w_2)}\right)
 e^{i\xi_1 w_1 + i\xi_2 w_2 + \frac{i}{3}(w_1^3+w_2^3)}.
\end{align}
\begin{align}
\label{I4asy}
 \tilde{I}_4
 &\sim
 (1-\a)^{2(u_1+u_2)} \frac{p_c(-\b_0)^3}{(1-p_c(-\b_0)^2)}
 (p_c(\b_0))^{(\xi_1+\xi_2)d(\b_0)N^{1/3}}
 \frac{1}{d(\b_0)^3 N} \notag\\
 &\quad
 e^{ -2N\l_0(\b_0)
 -\l_1(\b_0)c(\b_0)N^{2/3}(\t_1+\t_2)
 -\l_2(\b_0)c(\b_0)^2 N^{1/3}(\t_1^2+\t_2^2)
 -\l_3(\b_0)c(\b_0)^3 (\t_1^3+\t_2^3)+\xi_1\t_1+\xi_2\t_2} \notag\\ 
 &\quad
 \frac{-1}{4\pi^2}\int_{\text{Im}w_1=\eta_1} dw_1 
                 \int_{\text{Im}w_2=\eta_2} dw_2 
 \left(\tau_1-\tau_2+i(w_1-w_2)\right)
 e^{i\xi_1 w_1 + i\xi_2 w_2 + \frac{i}{3}(w_1^3+w_2^3)},
\end{align}
\begin{align}
\label{D4asy}
 D_4
 &\sim
 (1-\a)^{-2(u_1+u_2)} \frac{p_c(\b_0)^4}{(1-p_c(\b_0)^2)^3}
 (p_c(\b_0))^{-(\xi_1+\xi_2)d(\b_0)N^{1/3}}
 \frac{1}{d(\b_0)^3 N} \notag\\
 &\quad
 e^{ 2N\l_0(\b_0)
    +\l_1(\b_0)c(\b_0)N^{2/3}(\t_1+\t_2)
    +\l_2(\b_0)c(\b_0)^2 N^{1/3}(\t_1^2+\t_2^2)
    +\l_3(\b_0)c(\b_0)^3 (\t_1^3+\t_2^3)-\xi_1\t_1-\xi_2\t_2} \notag\\
 &\quad
 \frac{1}{4\pi^2}\int_{\text{Im}w_1=\eta_1} dw_1 
                 \int_{\text{Im}w_2=\eta_2} dw_2 
 \left(\tau_1-\tau_2-i(w_1-w_2)\right)
 e^{i\xi_1 w_1 + i\xi_2 w_2 + \frac{i}{3}(w_1^3+w_2^3)}.
\end{align}
Then discussions similar to ones below (\ref{Dasy}) lead
to (\ref{detKlim2}).
\qed

\subsubsection{Near the origin}
Next we show that the fluctuation near the origin is described 
by the process which gives the symplectic-unitary transition 
in random matrix theory \cite{FNH1999}. Namely, if we define the 
scaled height variable as (\ref{scaledHo}), one can show
\begin{theorem}
\begin{equation}
\label{detK4lim}
 \lim_{N\to\i} \P[H_N(\t_1) \leq s_1, \cdots , H_N(\t_m) \leq s_m]
 =
 \sqrt{\det(1+\K_4\G)},
\end{equation}
where $\G(\t_j,\xi)=-\chi_{(s_j,\i)}(\xi)$ ($j=1,2,\cdots,m$) and 
$\K_4$ is the $2\times 2$ matrix kernel, for which a representative 
element is given by 
\begin{equation}
 \K_4(\t_1,\xi_1;\t_2,\xi_2)
 =
 \begin{bmatrix}
  \S_4(\t_1,\xi_1;\t_2,\xi_2) & \D_4(\t_1,\xi_1;\t_2,\xi_2) \\
  \I_4(\t_1,\xi_1;\t_2,\xi_2) & \S_4(\t_2,\xi_2;\t_1,\xi_1) 
 \end{bmatrix},
\end{equation}
with the matrix elements being
\begin{align}
 \label{defS4lim}
 \S_4(\t_1,\xi_1;\t_2,\xi_2)
 &=
 \tilde{\S_4}(\t_1,\xi_1;\t_2,\xi_2)-\Phi_{\t_1,\t_2}(\xi_1,\xi_2), \\
 \tilde{\S_4}(\t_1,\xi_1;\t_2,\xi_2)
 &=
 \int_0^{\i} d\l e^{-\l(\t_1-\t_2)} \Ai(\xi_1+\l) \Ai(\xi_2+\l)
 -
 \frac12 \Ai(\xi_2) \int_{\xi_1}^{\i} d\l e^{-\l \t_1} \Ai(\xi_2-\l),\\
 \label{defI4im}
 \I_4(\t_1,\xi_1;\t_2,\xi_2)
 &=
 -\int_0^{\i} d\l e^{-\l \t_1} \Ai(\xi_1-\l) 
  \frac{d}{d \l} \left\{e^{-\l\t_2} \Ai(\xi_2-\l)\right\} \notag\\
 &\quad+\int_0^{\i} d\l e^{-\l \t_2} \Ai(\xi_2-\l) 
  \frac{d}{d \l} \left\{e^{-\l\t_1} \Ai(\xi_1-\l)\right\}\\ 
\label{defD4lim}
 \D_4(\t_1,\xi_1;\t_2,\xi_2)
 &=
 \frac{1}{4}\int_0^{\i} d\l e^{-\l\t_1} \Ai(\xi_1+\l) 
  \int_{\l}^{\i} dv e^{-v\t_2} \Ai(\xi_2+v) \notag\\
 &-
 \frac{1}{4}\int_0^{\i} d\l e^{-\l\t_2} \Ai(\xi_2+\l)
  \int_{\l}^{\i} dv \Ai(\xi_1+v) .
\end{align} 
Definition of $\Phi_{\t_1,\t_2}(\xi_1,\xi_2)$ is already given in 
(\ref{Phidef}). 
\end{theorem}

\vspace{3mm}\noindent
{\bf Remark.} 
For $\t=0$, we have
\begin{equation}
\label{F4}
 \lim_{N\to\i} \P[H_N(0) \leq s] = F_4(s),
\end{equation} 
where $F_4(s)$ is the GSE Tracy-Widom distribution 
\cite{TW1996}.  
Notice a notational difference in \cite{TW1996} and \cite{BR2001c}.
We follow the convention in the latter; our $F_4(s)$ is 
$F_4(s)=F_4^{\text{BR}}(s)=F_4^{\text{TW}}(\sqrt{2}s)$.
The result, (\ref{F4}), was shown in \cite{BR2001c};
it would also be possible to prove this by using
\begin{equation}
\label{MSE}
 \P [h(0,2N) \leq X]
 =
 \frac{1}{Z_N^{(4)}}
 \sum_{\begin{subarray}{c} 
        h\in\N^{\frac{N}{2}} \\ \text{max}\{h_j\}\leq X+N-1
       \end{subarray}}
 \prod_{1\leq i<j\leq \frac{N}{2}} (h_i-h_j)^2(h_i-h_j+1)(h_i-h_j-1)
 \prod_{i=1}^{\frac{N}{2}} q^{h_i},
\end{equation}
and the skew orthogonal polynomials techniques  
\cite{NW1991,NW1992a,NW1992b}.
The Meixner symplectic ensemble representation, (\ref{MSE}), 
was not given in \cite{Jo2000} but can be proved similarly if 
one notices that a symmetric $N\times N$ matrix with zero elements 
on diagonal and non-negative integer elements on off-diagonal 
is mapped to a semistandard Young tableau with all columns
of even length through Knuth correspondence \cite{Bu1974}.

\vspace{3mm}\noindent
{\bf Proof.} 
Derivation of (\ref{detK4lim}) is analogous to that of (\ref{detKlim}).
Using the integral representation of the Airy function, (\ref{Aiint}), 
we have, for $\tilde{\S}_4$, 
\begin{align}
 &\quad \tilde{\S}_4(\t_1,\xi_1;\t_2,\xi_2) \notag\\
 &=
 \frac{1}{4\pi^2}\int_{\text{Im}w_1=\eta_1} dw_1 
                 \int_{\text{Im}w_2=\eta_2} dw_2 
 \left(-\frac{1}{\tau_2-\tau_1+i(w_1+w_2)}-\frac{1}{2(\tau_1-iw_1)}\right)
 \notag\\
 &\quad
 e^{i\xi_1 w_1 + i\xi_2 w_2 + \frac{i}{3}(w_1^3+w_2^3)},
\end{align}
where $\eta_1+\eta_2+\t_1-\t_2>0$.
For $\I_4$, 
\begin{align}
 &\quad \I_4(\t_1,\xi_1;\t_2,\xi_2) \notag\\
 &=
 \frac{-1}{4\pi^2}\int_{\text{Im}w_1=\eta_1} dw_1 
                 \int_{\text{Im}w_2=\eta_2} dw_2 
 \frac{\tau_2+iw_2}{(\tau_1+\tau_2+i(w_1+w_2))}
 e^{i\xi_1 w_1 + i\xi_2 w_2 + \frac{i}{3}(w_1^3+w_2^3)} \notag\\
 &-  
 (\t_1,\xi_1 \leftrightarrow \t_2,\xi_2),
\end{align}
where $\t_1+\t_2-\eta_1-\eta_2>0$.
For $\D_4$, 
\begin{align}
 &\quad \D_4(\t_1,\xi_1;\t_2,\xi_2) \notag\\
 &=
 \frac{1}{4\pi^2}\int_{\text{Im}w_1=\eta_1} dw_1 
                 \int_{\text{Im}w_2=\eta_2} dw_2 
 \frac{1}{4(\tau_1-iw_1)(\tau_1+\tau_2-i(w_1+w_2))}
 e^{i\xi_1 w_1 + i\xi_2 w_2 + \frac{i}{3}(w_1^3+w_2^3)} \notag\\
 &-  
 (\t_1,\xi_1 \leftrightarrow \t_2,\xi_2),
\end{align}
where there is no additional condition on $\t_1,\t_2,\eta_1,\eta_1$. 
Then applying the same method as in 4.3, we can show
\begin{align}
 \label{S4lim}
 S_4(r_1,x_1;r_2,x_2)
 &\sim
 e^{(\t_2^3-\t_1^3)/3+\xi_1\t_1-\xi_2\t_2}dN^{1/3} 
 \S_4(\t_1,\xi_1;\t_2,\xi_2) , \\
 \label{I4lim}
 I_4(r_1,x_1;r_2,x_2)
 &\sim
 e^{(\t_1^3+\t_2^3)/3-\xi_1\t_1-\xi_2\t_2}d^2N^{2/3}
 \I_4(\t_1,\xi_1;\t_2,\xi_2) , \\
 \label{D4lim}
 D_4(r_1,x_1;r_2,x_2)
 &\sim
 e^{-(\t_1^3+\t_2^3)/3+\xi_1\t_1+\xi_2\t_2} 
 \D_4(\t_1,\xi_1;\t_2,\xi_2),
\end{align}
from which (\ref{detK4lim}) follows. 
\qed

\setcounter{equation}{0}
\setcounter{theorem}{1}
\section{Continuous Limit}
\label{continuous}
Let us take the limit $\a\to 0$, $N\to\i$ with $t=\a N,$ fixed.
In this limit time and space become continuous;
the time variable $t$ can take any positive value whereas
the scaled space variable is defined to be $v=\a r/2$.
The model is reduced to the standard PNG model in a half space 
($v\geq 0$) with an external source at the origin.
As in the discrete case, we extend the space to the whole space
($v\in\R$), putting the symmetry condition on the hight
with respect to the origin. 
First we have a flat substrate. At time 0, a nucleation of a height one
occurs at the origin. This step grows laterally in both directions 
with unit speed. 
Above this ground layer there occur other nucleations with rate two 
per unit length. The height of a nucleation is always one.
There is an external source at the origin with rate $\g$.
As in the discrete case, one can define the multi-layer version 
\cite{PS2002b}.

It is easy to take this limit at the level of the 
generating functions. For the orthogonal case, 
(\ref{Sgg1}),(\ref{Igg1}),(\ref{Dgg1}) become
\begin{align}
\label{Sgg1c}
 \tilde{\Sg}_1 
 &\sim
 e^{2(v_1-v_2)}
 e^{t(z_1-1/z_1)-v_1(z_1+1/z_1)}
 e^{t(1/z_2-z_2)+v_2(z_2+1/z_2)}
 \frac{z_1}{z_1-z_2} 
 \left\{ 1 + \frac{z_1-z_2}{(1+z_1)(z_2-1)} \right\}, \\
\label{Igg1c}
 \tilde{\Ig}_1 
 &\sim
 e^{-2(v_1+v_2)}
 e^{t(z_1-1/z_1)+v_1(z_1+1/z_1)}
 e^{t(1/z_2-z_2)+v_2(z_2+1/z_2)} 
 \frac{z_1}{z_1-z_2}
 \left\{ \frac12\frac{z_2+1}{z_2-1}-\frac12\frac{1+z_1}{1-z_1} \right\},
 \\
\label{Dgg1c}
 \tilde{\Dg}_1 
 &\sim
 e^{2(v_1+v_2)}
 e^{t(z_1-1/z_1)+v_1(z_1+1/z_1)}
 e^{t(1/z_2-z_2)+v_2(z_2+1/z_2)}
 \frac{z_1}{z_1-z_2} 
 \left\{ \frac12\frac{z_1-1}{z_1+1}-\frac12\frac{1-z_2}{1+z_2} \right\}.
\end{align}
Expressions for the symplectic case are given by
(\ref{Sgg1c}),(\ref{Igg1c}),(\ref{Dgg1c}) with the last factors 
replaced by the corresponding ones in (\ref{Sgg4}),(\ref{Igg4}),(\ref{Dgg4}).

The scaling limit can also be studied as in the discrete PNG model. 
First the thermodynamic shape is known to be
\begin{equation}
 h(v=\b_0 t,t) /t 
 \sim 
 2\sqrt{1-\b_0^2},
\end{equation}
where $0<\b_0<1$ is fixed \cite{KS1992}.
To consider the scaling limit in the bulk,
define the scaled height variable as
\begin{equation}
\label{scaledHcb}
 H_N(\t,\b_0) = \frac{h(v=\b_0 t+c(\b_0)t^{\frac23}\t,t)
                      -a(\b_0+\frac{c(\b_0)\t}{t^{1/3}}) t}
                     {d(\b_0) t^{\frac13}} ,
\end{equation}
where
\begin{align}
 a(\b) &=  2\sqrt{1-\b^2},\\
 d(\b) &= (1-\b^2)^{\frac16}, \\
 c(\b) &= (1-\b^2)^{\frac76}.
\end{align}
Then one can show 
\begin{proposition}
\label{cont_bulk}
For the scaled height variable defined 
by (\ref{scaledHcb}), (\ref{detKlim2}) holds.  
\end{proposition}
The proof of the Proposition \ref{cont_bulk} is parallel to that in \ref{bulk}.
The main difference is that the function $g_{\mu,\b}$ in (\ref{defg})
is replaced by a simpler
\begin{equation}
g_{\mu,\b}(z)
=
z-\frac{1}{z}-\b(z+\frac{1}{z})-\mu\log z,
\end{equation}
for which the double critical point is
\begin{align}
 \mu_c(\b) 
 &= 
 2\sqrt{1-\b^2}, \\
 \label{pcdef_c}
 p_c(\b)
 &=
 \sqrt{\frac{1+\b}{1-\b}}.
\end{align}

As for the fluctuation near the origin, 
if we define the scaled height variable as
\begin{equation}
\label{scaledHco}
 H_N(\t) = \frac{h(v=t^{\frac23}\t,t)-2 t}{t^{\frac13}}+\t^2,
\end{equation}
we have
\begin{proposition}
For the scaled height variable defined by (\ref{scaledHco}), one has
(\ref{detKlim}) for the orthogonal case and (\ref{detK4lim}) 
for the symplectic case.
\end{proposition}

\setcounter{equation}{0}
\setcounter{theorem}{1}
\section{Discussions}
\label{discussions}
In the preceding sections, we have given a detailed analysis 
of the height fluctuation of the model for two special values 
of the strength of the external source at the origin, $\g=1$
and $\g=0$.
Our results show that the fluctuation near the origin is
described by the orthogonal/symplectic to unitary transition
in random matrix theory.

This implies, in particular, that the height fluctuation of the 
PNG model at a single point near the origin is equivalent to 
that of the largest eigenvalue of the transition ensemble.
To check this, we performed Monte-Carlo simulations of the 
PNG model and the transition random matrix. In Figs. 4 and 5, 
we have shown the fluctuations of the height of the PNG model and the 
largest eigenvalue of the transition ensemble. We see an excellent 
agreement between them. 

Unfortunately properties of the transition ensemble have not been 
well studied compared to the ensembles with a specified symmetry.
For example, it seems difficult to numerically compute the 
probability distribution function of the largest eigenvalue .
This is sharply contrasted to the situation for the GUE/GOE/GSE, 
for which the Painlev\'e representation allows us to
plot the probability distribution function with high accuracy.
It would be desirable to study the transition ensemble to 
better understand the statistical properties of the PNG model.

Now we argue what happens for other values of $\g$.
For the fluctuation at the origin, the results have been 
obtained in \cite{BR2001c} where a strong universality is observed.
The GSE Tracy-Widom distribution describes the fluctuation 
not only for $\g=0$ but also for all values in $0\leq \g <1$.
The fluctuation becomes the Gaussian for all values in $\g>1$.
On the other hand, the GOE Tracy-Widom distribution appears 
only at $\g=1$.
The Gaussian fluctuation for $\g>1$ is stated only 
for the continuous model in \cite{BR2001c}, but is expected to 
persist for the discrete model as well.
In Fig. 6, typical shapes of the droplet are shown for 
three cases where $\g<1$, $\g=1$, $\g>1$. From these 
one should be able to guess that the shape looks similar 
for all values in $\g\leq 1$ but a cusp-like piece appears near
the origin when $\g$ becomes greater than unity.

This suggests that the fluctuation near the origin also has 
a similar universality.  
For all values in $0\leq \g <1$,
it is expected to be described by the symplectic-unitary 
transition. This is a conjecture because we do not know 
how to prove this at present, but is supported by 
a Monte-Carlo simulation. See Fig. 7, where a good 
agreement is observed.

For $\g>1$, we can also expect some universal behavior
for the fluctuation near the origin.
We have not found a compact formula for general $\g$,
but the situation becomes quite simple in the limiting case 
where $\a\to 0$ with $0<\g \a<1$ fixed.
In this limit, the nucleations in the bulk are so rare 
that the only those at the origin are important. 
Then the height at a position $r$ and at time $t$ would 
be almost the same as that at a position $0$ and at time $t-r$.
Therefore the height fluctuation at a single point is 
given by the Gaussian. If we set $r=\rho t$ with $0<\rho<1$
fixed, we have
\begin{equation}
\lim_{t\rightarrow\infty}
\P\left(\frac{h(r,t)-(1-\rho)a_G t}
{(1-\rho)^{\frac{1}{2}}d_G t^{\frac{1}{2}}}\leq s\right)
=
\frac{1}{\sqrt{2\pi}}\int_{-\i}^s e^{-\frac{\xi^2}{2}} d\xi
=
\text{erf}(s)
\end{equation}
where $a_G=\frac{\gamma\alpha}{1-\gamma\alpha}$ and 
$d_G=\frac{1}{1-\gamma\alpha}$.
A result of a Monte-Carlo simulation corresponding to this 
case is shown in Fig. 8.
In addition, since nucleations at the origin are independent 
for each time, the multi-point equal time height fluctuation 
would be described by the one-dimensional Brownian motion.
The situation would be somewhat more difficult for a smaller $\g$,
but we expect that the same Gaussian fluctuation is observed
in an appropriate limit.

\setcounter{equation}{0}
\setcounter{theorem}{1}
\section{Conclusion}
\label{conclusion}
In this paper, we have studied fluctuation properties of the 
one-dimensional polynuclear growth (PNG) model in a half space
with an external source at the origin.
We have mainly considered the model in a discrete space and time, 
but have also given results for the model in a continuous setting.
The results in the scaling limit are the same for both cases.

For two special values of the strength of the external source, $\g$,
we have performed a detailed analysis.
The $\g=1$ case corresponds to a critical point, 
which we call the orthogonal case.
The $\g=0$ case corresponds to the model without the external source, 
which we call the symplectic case.
The main results are (\ref{detKlim}) for the orthogonal case,
and (\ref{detK4lim}) for the symplectic case.
According to these the height fluctuation of the model near the 
origin is equivalent to those of the largest eigenvalue of the 
orthogonal/symplectic to unitary transition ensemble at soft 
edge in random matrix theory. 
We have also shown that the height fluctuation at bulk is 
described by the Airy process. 
For other values of $\g$, we have conjectured that 
the fluctuation is the symplectic-unitary type for
$0\leq \g<1$, whereas it is the Gaussian type for $\g>1$. 
Some Monte-Carlo simulation results are also presented
to confirm our results and conjectures.

\section*{Acknowledgment}
The authors would like to thank H. Spohn, M. Katori, T. Nagao, 
M. Pr\"ahofer and P. Ferrari for useful discussions and comments.
The first author is grateful to the Zentrum Mathematik, TU M\"unchen 
and in particular to H. Spohn for kind hospitality during his stay.
He also thanks T. Miyake and Y. Saiga for advice on numerical 
diagonalization of matrices.

%\bibliographystyle{unsrt}
%\bibliography{all}

\begin{thebibliography}{10}

\bibitem{KS1992}
J.~Krug and H.~Spohn.
\newblock Kinetic roughening of growing interfaces.
\newblock In C.~Godr\`eche, editor, {\em Solids far from Equilibrium: Growth,
  Morphology and Defects}, pages 479--582, 1992.

\bibitem{Me1998}
P.~Meakin.
\newblock {\em Fractals, scaling and growth far from equilibrium}.
\newblock Cambridge, 1998.

\bibitem{KPZ1986}
{M. Kardar, G. Parisi and Y. C. Zhang}.
\newblock Dynamic scaling of growing interfaces.
\newblock {\em Phys. Rev. Lett.}, 56:889--892, 1986.

\bibitem{GS1992}
{L.-H. Gwa and H. Spohn}.
\newblock Six-vertex model, roughened surfaces, and an asymmetric spin
  {Hamiltonian}.
\newblock {\em Phys. Rev. Lett.}, 68:725--728, 1992.

\bibitem{Ki1995}
D.~Kim.
\newblock Bethe ansatz solution for crossover scaling functions of the
  asymmetric {XXZ} chain and the {Kardar-Parisi-Zhang-type} growth model.
\newblock {\em Phys. Rev. E}, 52:3512--3524, 1995.

\bibitem{Jo2000}
K.~Johansson.
\newblock Shape fluctuations and random matrices.
\newblock {\em Commun. Math. Phys.}, 209:437--476, 2000.

\bibitem{PS2000a}
M.~Pr{\"a}hofer and H.~Spohn.
\newblock Universal distributions for growth processes in 1+1 dimensions and
  random matrices.
\newblock {\em Phys. Rev. Lett}, 84:4882--4885, 2000.

\bibitem{PS2000b}
M.~Pr{\"a}hofer and H.~Spohn.
\newblock Statistical self-similarity of one-dimensional growth processes.
\newblock {\em Physica A}, 279:342--352, 2000.

\bibitem{BR2000}
J.~Baik and E.~M. Rains.
\newblock Limiting distributions for a polynuclear growth model with external
  sources.
\newblock {\em J. Stat. Phys}, 100:523--541, 2000.

\bibitem{GTW2001}
{J. Gravner, C. A. Tracy and H. Widom}.
\newblock Limit theorems for height fluctuations in a class of discrete space
  and time growth models.
\newblock {\em J. Stat. Phys.}, 102:1085--1132, 2001.

\bibitem{GTW2002a}
{J. Gravner, C. A. Tracy and H. Widom}.
\newblock A growth model in a random environment.
\newblock {\em Ann. Probab.}, 30:1340--1369, 2002.

\bibitem{GTW2002b}
{J. Gravner, C. A. Tracy and H. Widom}.
\newblock Fluctuations in the composite regime of a disordered growth model.
\newblock {\em Commun. Math. Phys.}, 229:433--458, 2002.

\bibitem{PS2002a}
M.~Pr{\"a}hofer and H.~Spohn.
\newblock Current fluctuations for the totally asymmetric simple exclusion
  process.
\newblock In V.~Sidoravicius, editor, {\em In and out of equilibrium, vol. 51
  of \it Progress in Probability}, pages 185--204, 2002.

\bibitem{PS2002b}
M.~Pr{\"a}hofer and H.~Spohn.
\newblock Scale invariance of the {PNG} droplet and the {A}iry process.
\newblock {\em J. Stat. Phys.}, 108:1071--1106, 2002.

\bibitem{PS2002p}
M.~Pr{\"a}hofer and H.~Spohn.
\newblock Exact scaling functions for one-dimensional stationary {KPZ} growth.
  cond-mat/0212519.

\bibitem{BDJ1999}
{J. Baik, P. A. Deift and K. Johansson}.
\newblock On the distribution of the length of the longest increasing
  subsequence in a random permutation.
\newblock {\em J. Amer. Math. Soc.}, 12:1119--1178, 1999.

\bibitem{AD1999}
D.~Aldous and P.~Diaconis.
\newblock Longest increasing subsequences: from patience sorting to the
  {Baik-Deift-Johansson} theorem.
\newblock {\em Bull. Amer. Math. Soc.}, 36:413--432, 1999.

\bibitem{Jo2002}
K.~Johansson.
\newblock Non-intersecting paths, random tilings and random matrices.
\newblock {\em Probab. Theory Relat. Fields}, 123:225--280, 2002.

\bibitem{Jo2002p}
K.~Johansson.
\newblock Discrete polynuclear growth and determinantal processes.
  math.{PR}/0206208.

\bibitem{TW1994}
C.~A. Tracy and H.~Widom.
\newblock Level-spacing distributions and the {Airy} kernel.
\newblock {\em Commun. Math. Phys.}, 159:151--174, 1994.

\bibitem{Me1991}
M.~L. Mehta.
\newblock {\em Random Matrices}.
\newblock Academic, 2nd edition, 1991.

\bibitem{TW1996}
C.~A. Tracy and H.~Widom.
\newblock On orthogonal and symplectic matrix ensembles.
\newblock {\em Commun. Math. Phys.}, 177:727--754, 1996.

\bibitem{BR2001a}
J.~Baik and E.~M. Rains.
\newblock Algebraic aspects of increasing subsequences.
\newblock {\em Duke Math. J.}, 109:1--65, 2001.

\bibitem{BR2001b}
J.~Baik and E.~M. Rains.
\newblock The asymptotics of monotone subsequences of involutions.
\newblock {\em Duke Math. J.}, 109:205--281, 2001.

\bibitem{BR2001c}
J.~Baik and E.~M. Rains.
\newblock Symmetrized random permutations.
\newblock In P.~M. Bleher and A.~R. Its, editors, {\em Random Matrix Models and
  Their Applications}, pages 1--29, 2001.

\bibitem{Dy1962}
F.~J. Dyson.
\newblock A {B}rownian-motion model for the eigenvalues of a random matrix.
\newblock {\em J. Math. Phys}, 3:1191--1198, 1962.

\bibitem{Jo2001}
K.~Johansson.
\newblock Discrete orthogonal polynomial ensembles and the {P}lancherel
  measure.
\newblock {\em Annals of Math.}, 153:259--296, 2001.

\bibitem{FNH1999}
{P. J. Forrester, T. Nagao and G. Honner}.
\newblock Correlations for the orthogonal-unitary and symplectic transitions at
  the hard and soft edges.
\newblock {\em Nucl. Phys. B}, 553:601--643, 1999.

\bibitem{KM1959a}
S.~Karlin and L.~McGregor.
\newblock Coincidence properties of birth and death processes.
\newblock {\em Pacific J.}, 9:1109--1140, 1959.

\bibitem{KM1959b}
S.~Karlin and L.~McGregor.
\newblock Coincidence probabilities.
\newblock {\em Pacific J.}, 9:1141--1164, 1959.

\bibitem{Ba2000}
J.~Baik.
\newblock Random vicious walks and random matrices.
\newblock {\em Comm. Pure Appl. Math.}, 53:1385--1410, 2000.

\bibitem{NF2002}
{T. Nagao and P. J. Forrester}.
\newblock Vicious random walkers and a discretization of {G}aussian random
  matrix ensembles.
\newblock {\em Nucl. Phys. B}, 620:551--565, 2002.

\bibitem{NKT2003}
{T. Nagao, M. Katori and H. Tanemura}.
\newblock Dynamical correlations among vicious random walkers.
\newblock {\em Phys. Lett. A}, 307:29--35, 2003.

\bibitem{TW1998}
C.~A. Tracy and H.~Widom.
\newblock Correlation functions, cluster functions, and spacing distributions
  for random matrices.
\newblock {\em J. Stat. Phys.}, 92:809--835, 1998.

\bibitem{deB1955}
N.~G. de~Bruijn.
\newblock On some multiple integrals involving determinants.
\newblock {\em J. Indian Math. Soc.}, 19:133--151, 1955.

\bibitem{MW1973}
See for instance Ch. 9 of 
B.~M. McCoy and T.~T. Wu.
\newblock {\em The Two-Dimensional Ising Model}.
\newblock Harvard University Press, 1973.

\bibitem{Ro1981}
H.~Rost.
\newblock Non-equilibrium behavior of a many particle process: Density profile
  and local equilibria.
\newblock {\em Zeitschrift f. Warsch. Verw. Gebiete}, 58:41--53, 1981.

\bibitem{Mac1994}
A.~M.~S. Mac\^edo.
\newblock Universal parametric correlations at the soft edge of spectrum of
  random matrix ensembles.
\newblock {\em Europhys. Lett.}, 26:641--646, 1994.

\bibitem{NF1998}
{T. Nagao and P. J. Forrester}.
\newblock Multilevel dynamical correlation functions for {D}yson's {B}rownian
  motion model of random matrices.
\newblock {\em Phys. Lett. A}, 247:42--46, 1998.

\bibitem{TW2003p}
C.~A. Tracy and H.~Widom.
\newblock System of differential equations for the Airy process.
  {math.PR/0302033}.

\bibitem{AvM2003p}
M.~Adler and P.~van Moerbeke.
\newblock {A PDE for the joint distributions of the Airy process}.
  {math.PR/0302329}.

\bibitem{Ka2003u}
M.~Katori and H.~Tanemura,
\newblock in preparation.

\bibitem{NW1991}
T.~Nagao and M.~Wadati.
\newblock Correlation functions of random matrix ensembles related to classical
  orthogonal polynomials.
\newblock {\em J. Phys. Soc. Jpn}, 60:3298--3322, 1991.

\bibitem{NW1992a}
T.~Nagao and M.~Wadati.
\newblock Correlation functions of random matrix ensembles related to classical
  orthogonal polynomials ii.
\newblock {\em J. Phys. Soc. Jpn}, 61:78--88, 1992.

\bibitem{NW1992b}
T.~Nagao and M.~Wadati.
\newblock Correlation functions of random matrix ensembles related to classical
  orthogonal polynomials iii.
\newblock {\em J. Phys. Soc. Jpn}, 61:1910--1918, 1992.

\bibitem{Bu1974}
W.~H. Burge.
\newblock Four correspondences between graphs and generalized Young tableaux.
\newblock {\em J. Comb. Th. (A)}, 17:12--30, 1974.

\end{thebibliography}

\newpage
%%%%%%%%%%%%%%%%%%%%%%%%%%%%%%%%%
%%% Figure Captions           %%%
%%%%%%%%%%%%%%%%%%%%%%%%%%%%%%%%%
\begin{large}
\noindent
Figure Captions
\end{large}

%%% Fig. 1 %%%%%%%%%%%%%%%%%%%%%%
\vspace{10mm}
\noindent
Fig. 1: 
Dynamical rules of the discrete PNG model. 
(a) 
At time $t$, a nucleation of a height $k$ occurs at site $r$ 
with probability $(1-q)q^k$ when $r\neq 0$ and 
$(1-\g\sqrt{q})(\g \sqrt{q})^k$ when $r=0$.
Once the nucleation occurs, it grows laterally toward right and
left with unit speed.
(b) 
When two steps collide, the one with higher height 
swallow the one with lower height.
In the multi-layer version, a nucleation occurs in the  
lower layer at the position above which the collision occurs
in the upper layer. The height is equal to the height of the 
swallowed region in the upper layer.

%%% Fig. 2 %%%%%%%%%%%%%%%%%%%%%%
\vspace{8mm}
\noindent
Fig. 2: 
An example of the multi-layer PNG heights for the $\g=1$ case
for $q=0.25$ at (a) $t=6$ (b) $t=100$. 
It looks symmetric with respect to $r=1/2$.
The limiting shape is also shown as a dotted line for (b).

%%% Fig. 3 %%%%%%%%%%%%%%%%%%%%%%
\vspace{8mm}
\noindent
Fig. 3: 
An example of the multi-layer PNG heights for the $\g=0$ case
for $q=0.25$ at $t=100$. 
Each neighboring two heights form a pair at $r=0$.
The limiting shape is also shown as a dotted line.

%%% Fig. 4 %%%%%%%%%%%%%%%%%%%%%%
\vspace{8mm}
\noindent
Fig. 4:
Probability distributions of the scaled height of the PNG model 
for $\g=1$ and the scaled largest eigenvalue of the 
orthogonal-unitary transition random matrix ensemble. 
For the PNG model, the parameters are for $q=0.25$, $t=2000$, 
and (a) $r=0$ (b) $r=330$ (c) $r=1000$ (each 10000 samples).
The first two cases correspond to (a) $\t=0$ (b) $\t=1$, 
in which the scaling (\ref{scaledHo}) is used. 
The case (c) is better fitted to the bulk scaling limit;
the scaling (\ref{scaledH}) is used.
In the figures, they are represented as circles. 
For the transition random matrix ensemble, the data are for $N=500$,
(a) $\t=0$ (b) $\t=1$ (c) $\t=10$ (each 1000 samples). 
In the figures, they are represented as $+$.
For (a)(resp. (c)), the GOE (resp. GUE) Tracy-Widom distribution 
is also shown as a solid line.

%%% Fig. 5 %%%%%%%%%%%%%%%%%%%%%%
\vspace{8mm}
\noindent
Fig. 5:
Probability distributions of the scaled height of the PNG model 
for $\g=0$ and the scaled largest eigenvalue of the 
symplectic-unitary transition random matrix ensemble. 
For the PNG model, the parameters are for $q=0.25$, $t=2000$, 
and (a) $r=0$ (b) $r=330$ (c) $r=1000$ (each 10000 samples).
The first two cases correspond to (a) $\t=0$ (b) $\t=1$, 
in which the scaling (\ref{scaledHo}) is used. 
The case (c) is better fitted to the bulk scaling limit;
the scaling (\ref{scaledH}) is used.
In the figures, they are represented as circles. 
For the transition random matrix ensemble, the data are for $N=200$,
(a) $\t=0$ (b) $\t=1$ (c) $\t=10$ (each 10000 samples). 
In the figures, they are represented as $+$.
For (a)(resp. (c)), the GSE (resp. GUE) Tracy-Widom distribution 
is also shown as a solid line.

%%% Fig. 6 %%%%%%%%%%%%%%%%%%%%%%%%%%%
\vspace{8mm} 
\noindent
Fig. 6:
Typical droplet shapes for three values of 
$\gamma=0.8,1.0$ and $1.2$, which are fairly close to 
the critical value $\g=1$.
The model parameters are taken to be $t=1000, q=0.25$.
Notice that the appearance of the shape changes drastically when
$\gamma$ gets greater than one.

%%% Fig. 7 %%%%%%%%%%%%%%%%%%%%%%
\vspace{8mm}
\noindent
Fig. 7:
Probability distributions of the scaled height of the PNG model 
for $\g=0.5$ and the scaled largest eigenvalue of the 
symplectic-unitary transition random matrix ensemble. 
Parameters are the same as those for Fig. 5.

%%% Fig. 8 %%%%%%%%%%%%%%%%%%%%%%
\vspace{8mm}
\noindent
Fig. 8:
Probability distributions of the scaled height of the PNG model 
for $q=0.0001, \g=50$ and the Gaussian distribution (solid line).
For the PNG model, the parameters are $t=2000$ 
and $r=0$ ($+$), $r=330$ ($\circ$), $r=1000$ ($\bullet$)
(each 10000 samples).

%%%%%%%%%%%%%%%%%%%%%%%%%%%%%%%%%
%%% Figures                   %%%
%%%%%%%%%%%%%%%%%%%%%%%%%%%%%%%%%
%%%%%%%%%%% Fig. 1: discrete PNG  %%%%%%%%%
\newpage
\renewcommand{\thepage}{Figure 1}
{\unitlength=0.5cm
\begin{picture}(25,6)
\put(-2,5.5){(a)}
\put(0,1.1){\line(1,0){10}}
\multiput(2,1)(2,0){4}{\line(0,1){0.2}}
\thicklines
\put(0,1.1){\line(1,0){10}}
\thinlines
\put(11,1.1){\vector(1,0){3}}
\put(15,1.1){\line(1,0){10}}
\multiput(17,1)(2,0){4}{\line(0,1){0.2}}
\thicklines
\multiput(15,1.1)(6,0){2}{\line(1,0){4}}
\multiput(19,1.1)(2,0){2}{\line(0,1){4}}
\put(19,5.1){\line(1,0){2}}
\thinlines
\multiput(21,5.05)(0.2,0){5}{.}
\put(21.6,3.6){\vector(0,1){1.5}}
\put(21.6,2.6){\vector(0,-1){1.5}}
\put(21.5,2.8){$k$}
\put(18,2.6){\vector(-1,0){1.3}}
\put(23,2.6){\vector(1,0){1.3}}
\multiput(1.3,0.3)(14.9,0){2}{$r\!-\!1$}
\multiput(3.8,0.3)(15,0){2}{$r$}
\multiput(5.3,0.3)(14.9,0){2}{$r\!+\!1$}
\multiput(7.3,0.3)(15,0){2}{$r\!+\!2$}
\put(18.8,4.9){$\bullet$}
\put(20.8,1.0){$\bullet$}
\end{picture}}

{\unitlength=0.5cm
\begin{picture}(25,10)
\put(-2,7.3){(b)}
\put(0,1.1){\line(1,0){10}}
\multiput(0.5,1)(1.8,0){6}{\line(0,1){0.2}}
\put(11,1.1){\vector(1,0){3}}
\put(15,1.1){\line(1,0){10}}
\multiput(15.5,1)(1.8,0){6}{\line(0,1){0.2}}
\multiput(-0.2,0.3)(15,0){2}{$r\!-\!\!2$}
\multiput(1.7,0.3)(15,0){2}{$r\!-\!\!1$}
\multiput(4.0,0.3)(15,0){2}{$r$}
\multiput(5.1,0.3)(15,0){2}{$r\!+\!1$}
\multiput(7.0,0.3)(15,0){2}{$r\!+\!2$}
\multiput(8.8,0.3)(15,0){2}{$r\!+\!3$}
\thicklines
\put(0,1.6){\line(1,0){10}}
\put(0,3.4){\line(1,0){2.3}}
\put(2.3,5.2){\line(1,0){1.8}}
\put(4.1,3.4){\line(1,0){1.8}}
\put(5.9,7.0){\line(1,0){1.8}}
\put(7.7,3.4){\line(1,0){2.3}}
\multiput(2.3,3.4)(1.8,0){2}{\line(0,1){1.8}}
\multiput(5.9,3.4)(1.8,0){2}{\line(0,1){3.6}}
\put(15,1.6){\line(1,0){4.1}}
\put(19.1,3.4){\line(1,0){1.8}}
\put(20.9,1.6){\line(1,0){4.1}}
\multiput(19.1,1.6)(1.8,0){2}{\line(0,1){1.8}}
\put(15,3.4){\line(1,0){0.5}}
\put(15.5,5.2){\line(1,0){3.6}}
\put(19.1,7.0){\line(1,0){5.4}}
\put(24.5,3.4){\line(1,0){0.5}}
\put(15.5,3.4){\line(0,1){1.8}}
\put(19.1,5.2){\line(0,1){1.8}}
\put(24.5,3.4){\line(0,1){3.6}}
\put(2.1,5.0){$\bullet$}
\put(3.9,3.2){$\bullet$}
\put(5.7,6.8){$\bullet$}
\put(7.5,3.2){$\bullet$}
\put(15.3,5.0){$\bullet$}
\put(18.9,6.8){$\bullet$}
\put(18.9,3.2){$\bullet$}
\put(20.7,1.4){$\bullet$}
\put(24.3,3.2){$\bullet$}
\end{picture}}

%%%%%%%%%%% Fig. 2: multi PNG orthogonal %%%%%%%%%
\newpage
\renewcommand{\thepage}{Figure 2}
\begin{picture}(400,600)
\put(90,300){\includegraphics{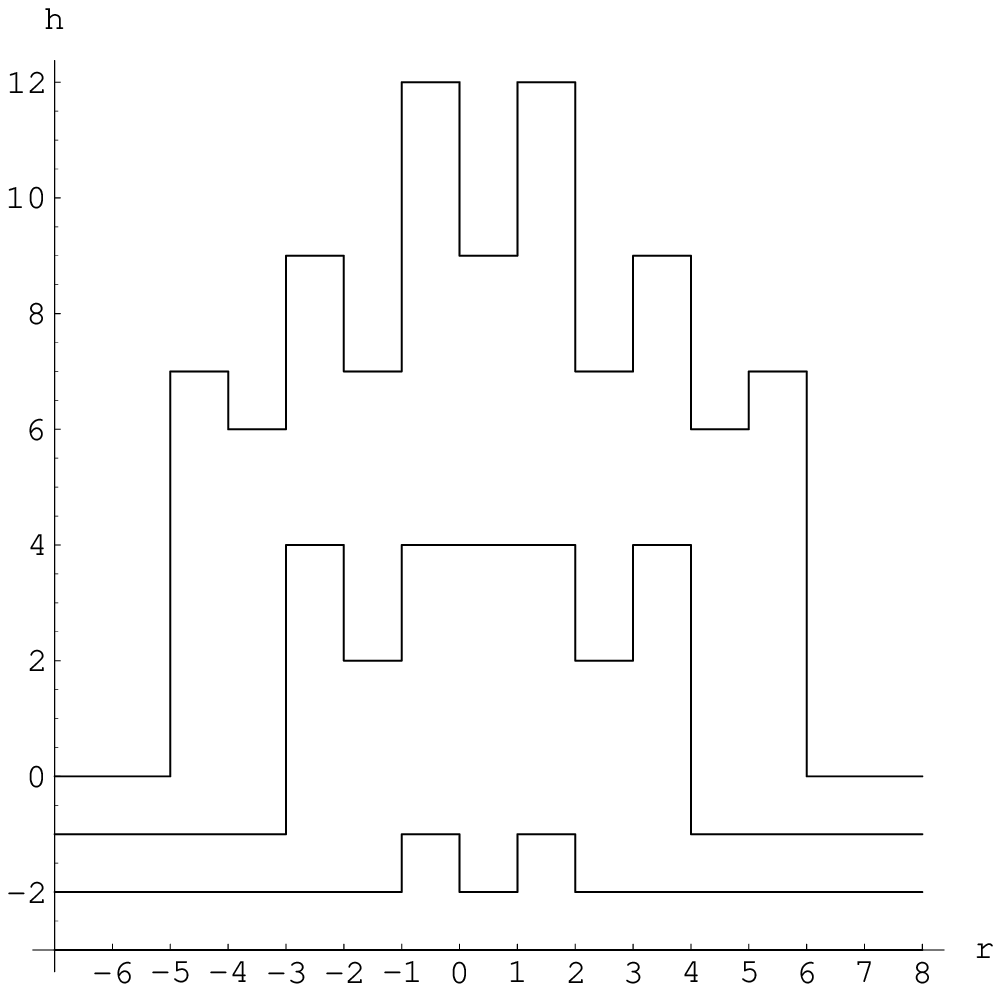}}
\put(50,0){\includegraphics{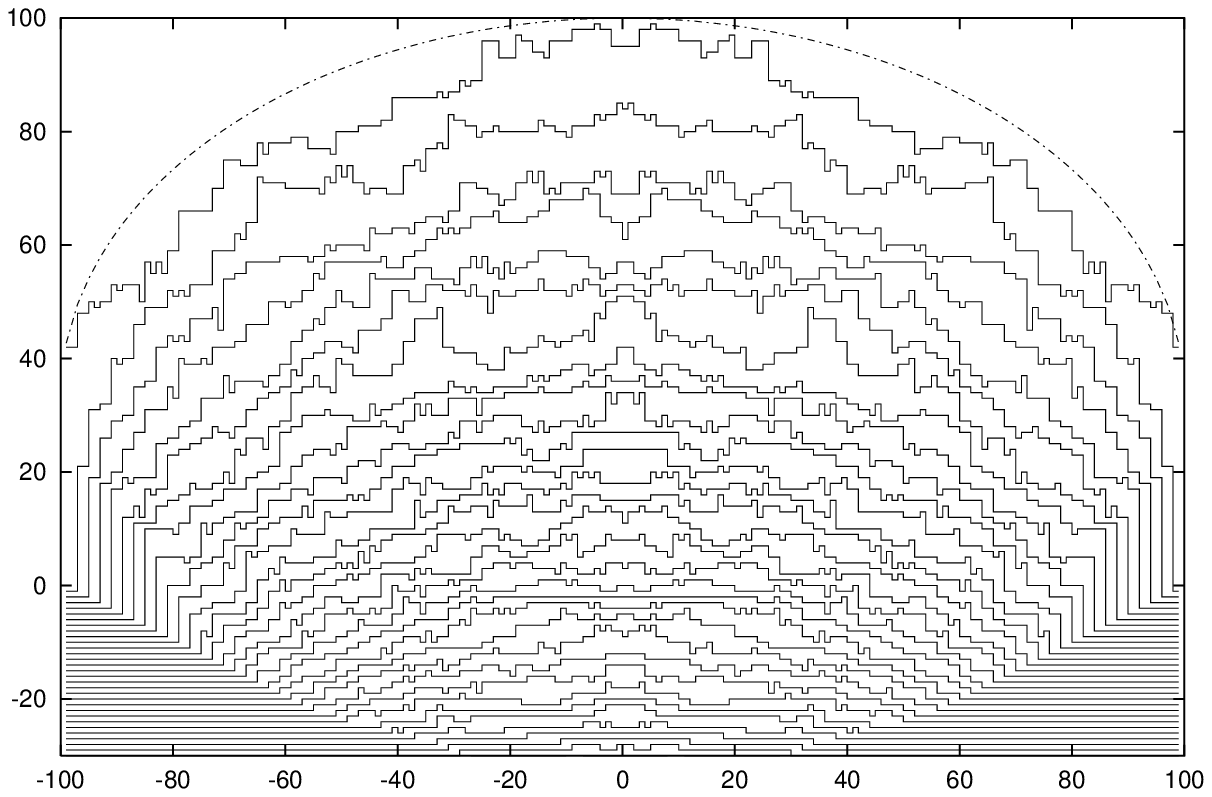}}
\put(20,580){(a)}
\put(20,260){(b)}
\put(43,228){$h$}
\put(400,-10){$r$}
\end{picture}

%%%%%%%%%%% Fig. 3: multi PNG symplectic %%%%%%%%%
\newpage
\renewcommand{\thepage}{Figure 3}
\begin{picture}(400,300)
%\put(80,300){\includegraphics{png_s.eps}}
\put(50,0){\includegraphics{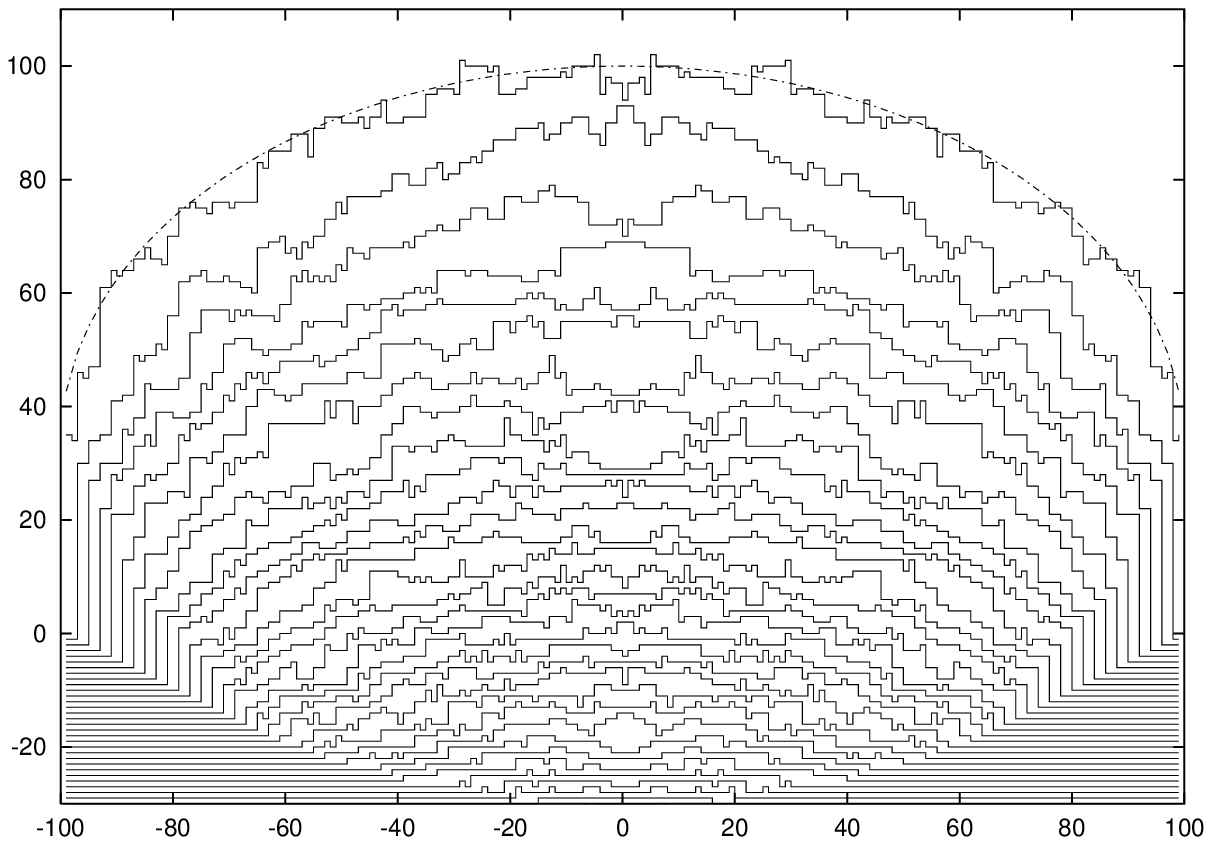}}
\put(43,240){$h$}
\put(400,-10){$r$}
\end{picture}

%%%%%%%%%%% Fig. 4: Fluctuation for \g = 1 %%%%%%%%%
\newpage
\renewcommand{\thepage}{Figure 4}
\begin{picture}(400,600)
\psfrag{Pt0}{\hspace*{-10mm}\large (a) $\tau=0$}
\psfrag{Pt1}{\hspace*{-10mm}\large (b) $\tau=1$}
\psfrag{Pr0}{\hspace*{-9mm} \large  (c) bulk}

\put(50,500){\includegraphics{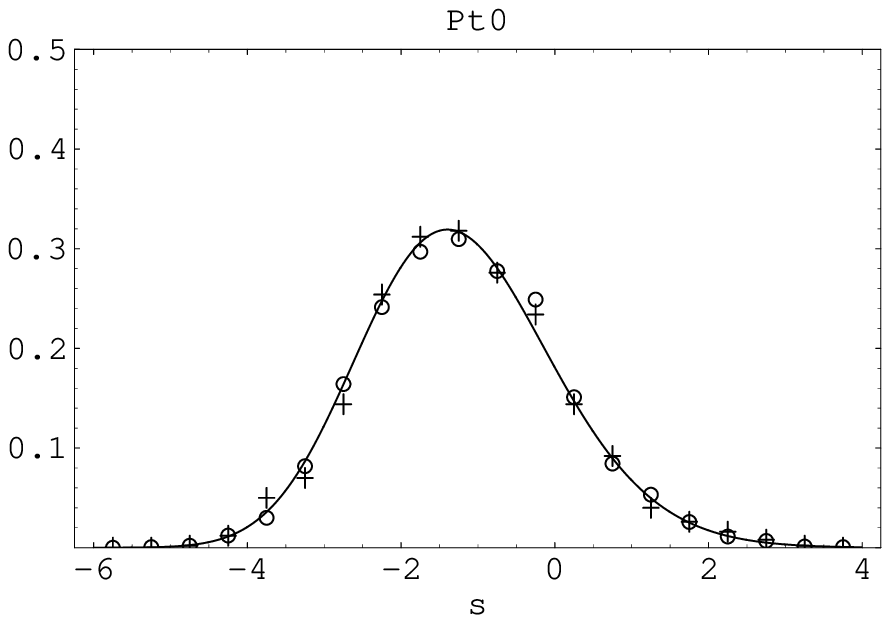}}
\put(50,300){\includegraphics{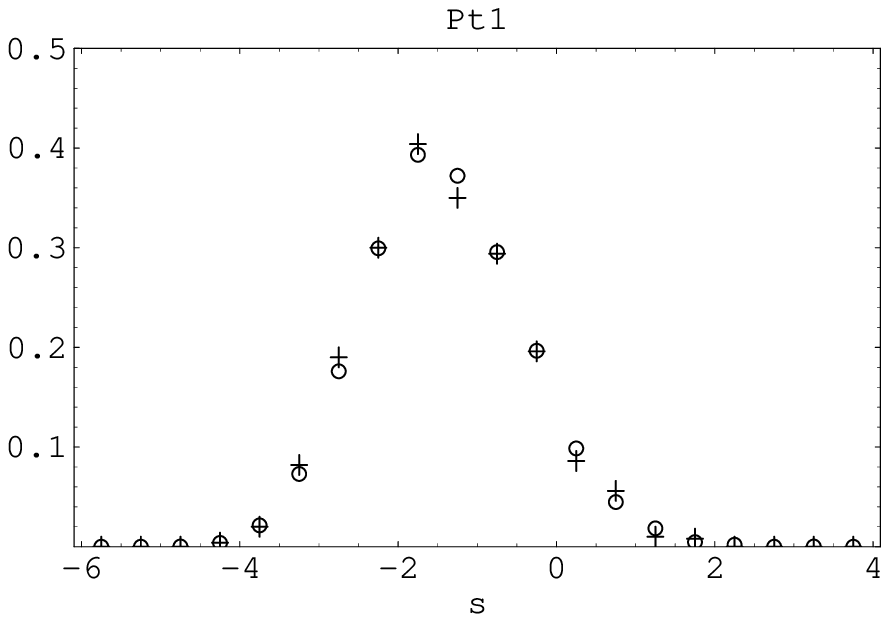}}
\put(50,100){\includegraphics{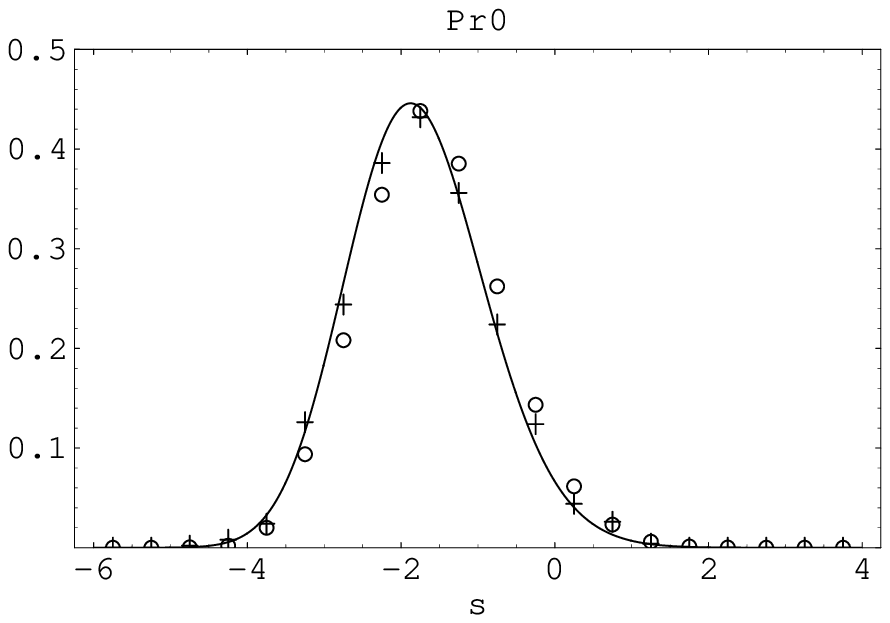}}

\end{picture}

%%%%%%%%%%% Fig. 5: Fluctuation for \g = 0 %%%%%%%%%
\newpage
\renewcommand{\thepage}{Figure 5}
\begin{picture}(400,600)
\psfrag{Pt0}{\hspace*{-10mm}\large (a) $\tau=0$}
\psfrag{Pt1}{\hspace*{-10mm}\large (b) $\tau=1$}
\psfrag{Pr0}{\hspace*{-9mm}\large (c) bulk}
\put(50,500){\includegraphics{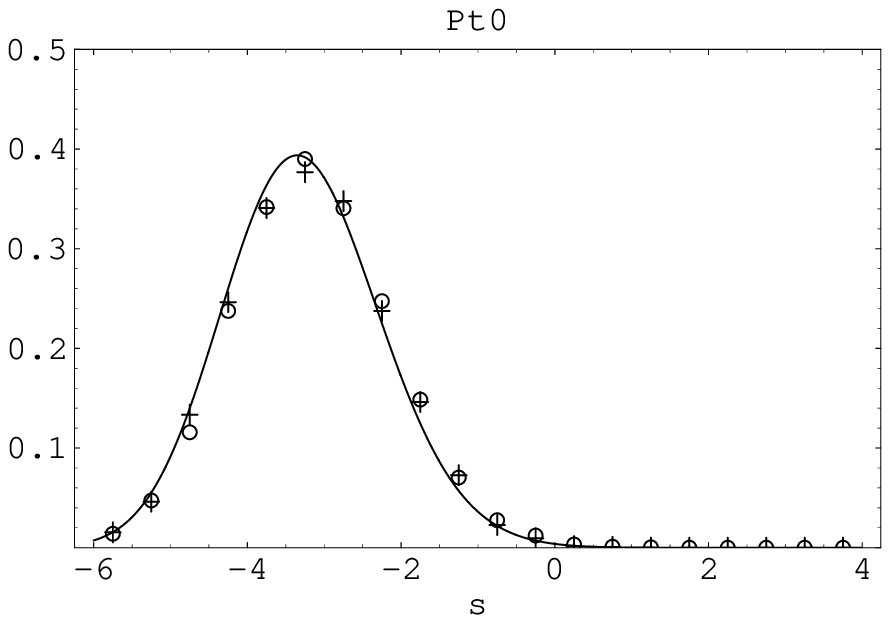}}
\put(50,300){\includegraphics{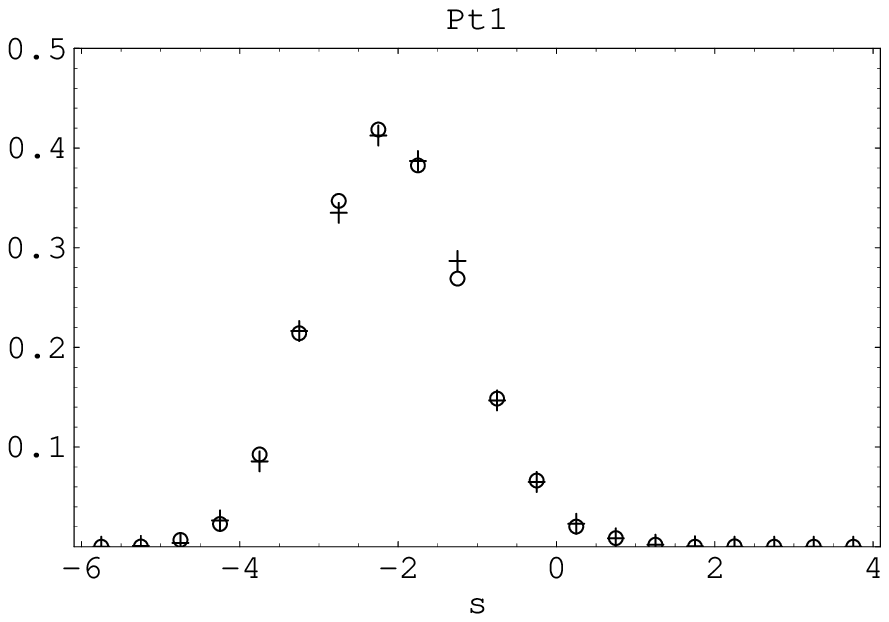}}
\put(50,100){\includegraphics{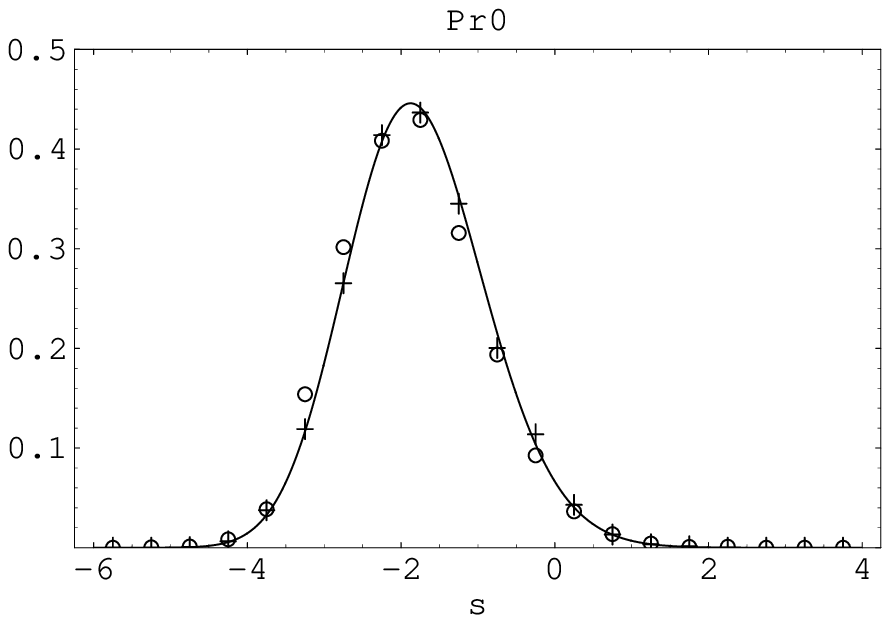}}
\end{picture}

%%%%%%%%%%% Fig. 6: Droplet Shapes near the Critical point%%%%%%%%%
\newpage
\renewcommand{\thepage}{Figure 6}
\begin{picture}(400,300)
\psfrag{gamma=0.8}{\scriptsize{$\gamma =0.8$}}
\psfrag{gamma=1.0}{\scriptsize{$\gamma=1.0$}}
\psfrag{gamma=1.2}{\scriptsize{$\gamma=1.2$}}
\put(50,100){\includegraphics{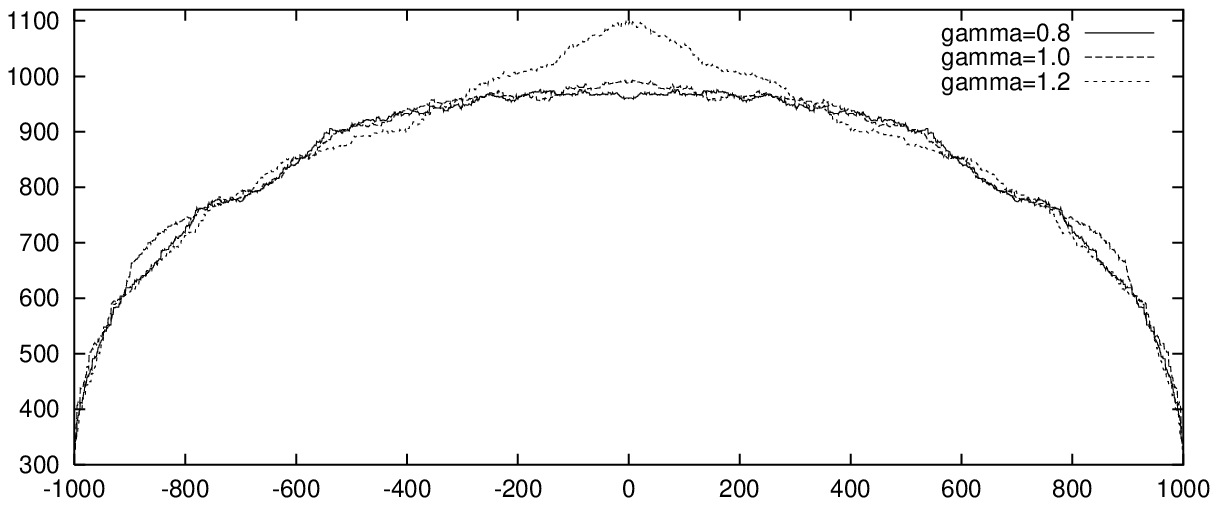}}
\end{picture}

%%%%%%%%%%% Fig. 7: Fluctuation for \g = 0.5 %%%%%%%%%
\newpage
\renewcommand{\thepage}{Figure 7}
\begin{picture}(400,600)
\psfrag{Pt0}{\hspace*{-10mm}\large (a) $\tau=0$}
\psfrag{Pt1}{\hspace*{-10mm}\large (b) $\tau=1$}
\psfrag{Pr0}{\hspace*{-9mm} \large (c) bulk}
\put(50,500){\includegraphics{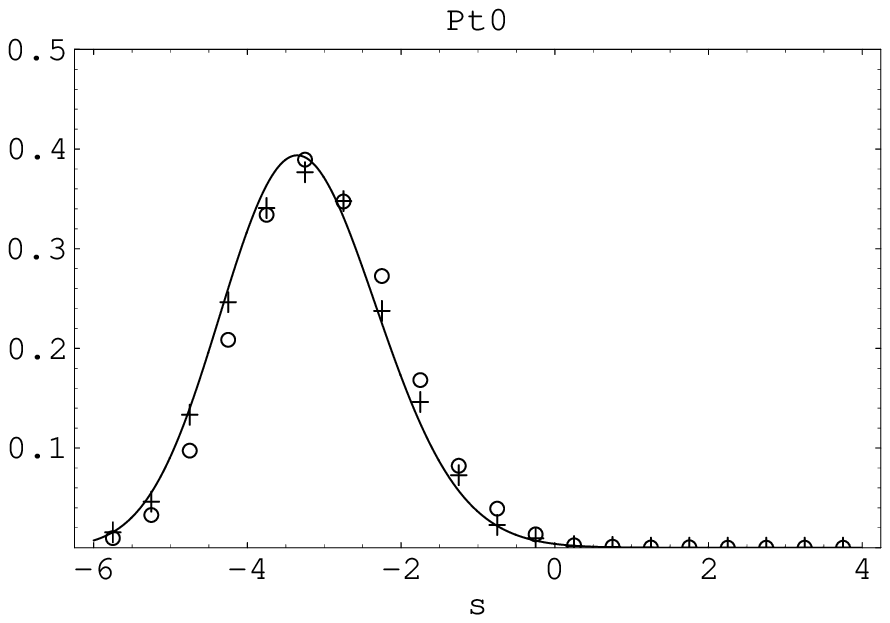}}
\put(50,300){\includegraphics{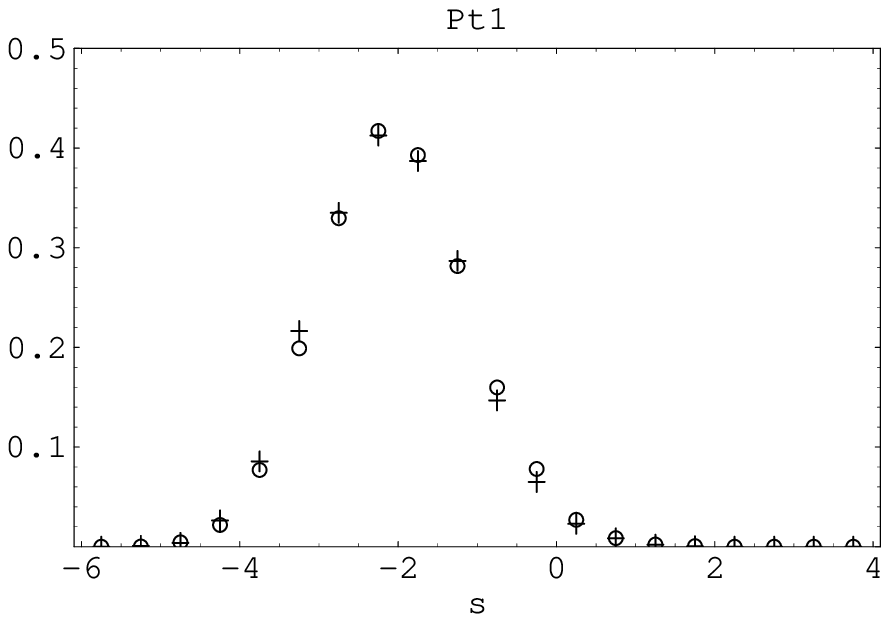}}
\put(50,100){\includegraphics{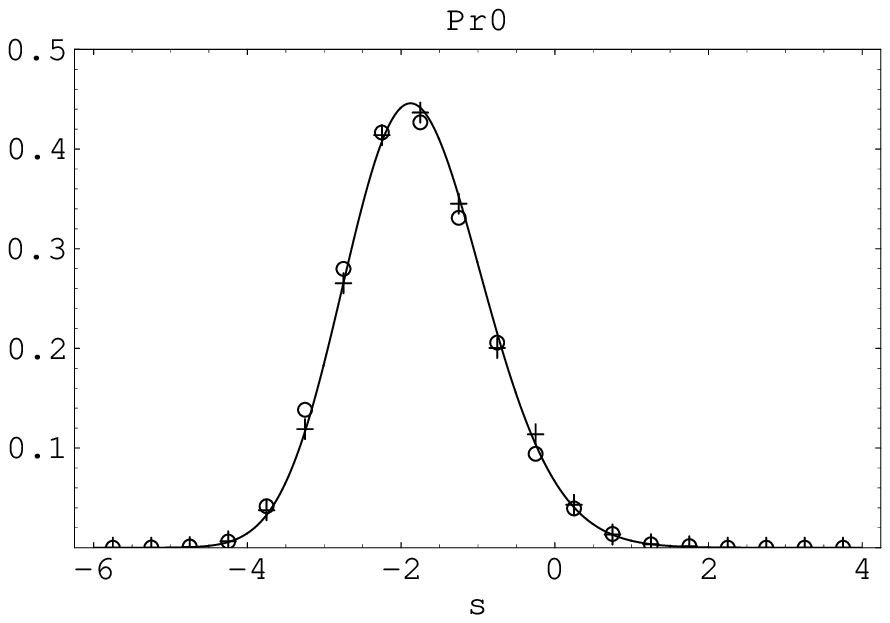}}
\end{picture}

%%%%%%%%%%% Fig. 8: Fluctuation for gaussian  %%%%%%%%%
\newpage
\renewcommand{\thepage}{Figure 8}
\begin{picture}(400,300)
\psfrag{Pg}{\hspace*{-22mm}\large Probability Densities}
\put(50,100){\includegraphics{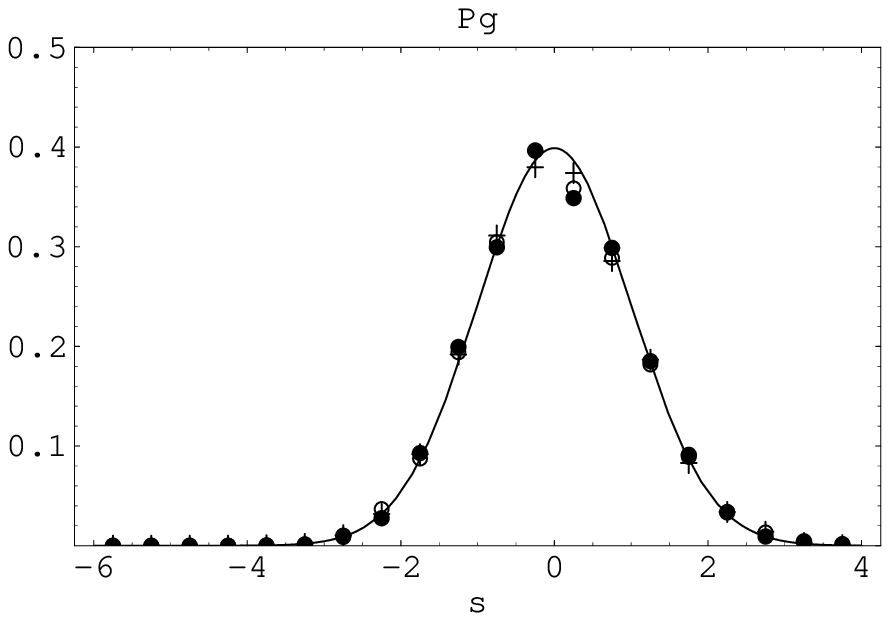}}
\end{picture}

\end{document}